\documentstyle[prd,aps,epsfig,eqsecnum,amssymb,amsmath,floats]{revtex}
\tighten
\begin{document}
\preprint{}
\draft
\title%%%%%%%%%%%%%%%%%%%%%%%%%%%%%%%%%%%%%%%%%%%%%%%%%%%%%%%%%%%%%%
{The \protect{$K_{\ell 3}$} Form Factors and \\ Atmospheric
                    Neutrino Flavor Ratio at High Energies}
\author{ V.~A.~Naumov\and${}^{1,2,3}$
            T.~S.~Sinegovskaya\and${}^2$ S.~I.~Sinegovsky\and${}^2$}
\address{ ${}^1$Laboratory of Theoretical Physics, Irkutsk State
               University, Irkutsk 664003, Russia                \\
         ${}^2$Department of Theoretical Physics, Physics Faculty,
               Irkutsk State University, Irkutsk 664003, Russia  \\
         ${}^3$Istituto Nazionale di Fisica Nucleare, Sezione di
               Firenze, Firenze 50125, Italy                      }
\maketitle
\date{\today}

\begin{abstract}
  We calculated the differential and total rates
  for the semileptonic decays of $K_L^0$ and $K^\pm$ mesons taking
  into account a linear $q^2$ dependence of the $K_{\ell3}$ form
  factors. As a case in point, we included these rates into the
  calculation of the atmospheric neutrino flux at energies 1 to
  100~TeV. The calculated neutrino spectra are between the earlier
  predictions while the neutrino flavor ratio, $R$, is somewhat
  affected by the $K_{\ell3}$ form factors. The $R$ proves to be very
  sensitive to the contribution of neutrinos from decay of charmed
  particles, providing an additional method to test the charm
  production models in future experiments with large-volume neutrino
  telescopes.

\end{abstract}

%\newpage
\widetext

\section{Introduction} \label{sec:I}

Atmospheric neutrinos (AN) come from the decays of unstable particles
generated in the collisions of primary and secondary cosmic rays with
air nuclei. Up to very high energies, the dominant contribution is
due to decays of charged pions, charged and long-lived neutral kaons.
The AN from these sources have come to be known as ``$\pi,K$'' or
conventional neutrinos. As energy increases, semileptonic decays of
charmed hadrons (mainly $D^\pm$, $D^0$, $\overline{D}{}^0$, $D_s^\pm$
mesons and $\Lambda_c^+$ hyperons) should become important. These AN
are usually called prompt neutrinos (PN). The borderline energy
between the domination regions of conventional and prompt neutrinos
is a long-standing question. Taking into account the available data on
cosmic-ray muons~\cite{Bugaev93}, it is safe to say that the PN fraction
in the vertical (horizontal) muon neutrino flux is negligible at energies
below $E_\nu^b \sim 1~TeV$ ($\sim 10~TeV$); for the electron
neutrino flux, the corresponding borders are roughly an order of
magnitude less. But it is not excluded that $E_\nu^b$ may be
increased by 10 to 100 times.

The AN flux with energies above 1~TeV represents an unavoidable
background for many of the astrophysical experiments with the
full-size underwater/ice neutrino telescopes%
\footnote{Among them the detection of neutrinos from the (quasi)diffuse
          neutrino backgrounds, like pregalactic neutrinos, neutrinos
          from the bright phase of galaxy evolution and from active
          galactic nuclei.}.
At the same time, the AN flux is a natural tool for studying neutrino
interactions with matter
%%%%%%%%%%%%%%%%%%%%%%%%%%%%%%%%%%%%%%%%%%%%%%%%%%%%%%%%%%%%%%%%%%%%%
%% \footnote{Measurements of the cross sections for $\nu_\ell N$
%%          ($\overline{\nu}_\ell N$) charged-current interactions at
%%          $\sqrt{s} \sim m_W$ ($E_\nu \sim 3.4~TeV$) provide an
%%          important test for the standard model. With modern
%%          accelerators, the neutrino interactions are studied at
%%          energies up to several hundreds of GeV (besides the single
%%          very high energy HERA data point extracted from the
%%          $e p\rightarrow\nu_e X$ cross section), whereas deep
%%          underwater experiments with AN will enable to enlarge the
%%          region of neutrino energies, up to a few tens of TeV. It
%%          is hoped that future ``KM3'' telescopes will be able to
%%          study the production of the standard vector $q\overline{q}$
%%          resonances ($\rho$, $D_s^*$ and possibly $\overline{t}b$)
%%          and the resonant $W^-$ production ($E_\nu^{\sy{res}} =
%%          m_W^2/(2m_e \simeq 6.3~PeV$) in $\overline{\nu}_e e^-$
%%          annihilation as well as hypothetical nonstandard
%%          interactions of neutrinos like interactions induced by
%%          off-diagonal neutral currents, the charged-current
%%          processes with production of supersymmetric particles or
%%          with an exchange of light leptoquarks and so forth.}
%%%%%%%%%%%%%%%%%%%%%%%%%%%%%%%%%%%%%%%%%%%%%%%%%%%%%%%%%%%%%%%%%%%%%
and, along with the atmospheric muon flux, it provides a way of
testing the inputs of nuclear cascade models that is parameters of
the primary cosmic-ray flux and cross sections of hadron-nucleus
interactions at energies beyond the reach of accelerator experiments.
In particular, the AN flux measurements have much potential for
yielding information about the mechanism of charm production.

In both cases -- to correct for the AN background and to use the AN
flux as an additional tool of particle physics -- it is necessary to
know with confidence the conventional AN flux.
The relevant calculations have been performed by many authors (see
\cite{Volkova80,Mitsui86,Butkevich89,Lipari93,Agrawal96,Bugaev89,%
Thunman95} and references therein).
The discrepancy between the different predictions for the $\pi,K$
neutrino flux above 1~TeV ranges up to 25--35\,\% for
$\nu_\mu+\overline{\nu}_\mu$ and to 60--70\,\% for
$\nu_e+\overline{\nu}_e$ (depending upon the zenith angle).
Unfortunately, these numbers are not representative of the upper
limits for the calculation uncertainty. For the most part this
is due to the incompleteness in the current knowledge of the primary
spectrum and composition as well as mechanisms of $\pi$ and, to a
greater extent, $K$ meson production at high energies
(see \cite{Agrawal96} for a detailed discussion).
A sizable part of the discrepancy is caused by different
approximations and simplifications employed in the calculations.

Predictions for the PN contribution vary by a few orders of
magnitude (see \cite{Bugaev89,Thunman95,Battistoni96} and references
therein). Here, the basic challenge is of course in the mechanism of
charm hadroproduction. An additional (though not so drastic) source
of uncertainty has to do with the differential rates of inclusive
semileptonic decays of charmed hadrons. The theory of charm production
and decay is still far from completion and the corresponding
accelerator data are rather meagre. However, when correlating the
predictions of different models with each other and with experiment,
the neglecting some comparatively small effects may introduce
additional systematic errors and should be avoided.

Some relative characteristics of the AN flux, like the zenith-angle
distribution and the flavor ratio,
\[
R=\frac{\nu_\mu+\overline{\nu}_\mu}{\nu_e+\overline{\nu}_e},
\]
prove to be less sensitive to the uncertainties of the conventional
AN flux predictions and, on the other hand, they are very dependent
of the PN contribution. The latter is owing to the salient and
model-independent features of the PN flux. First, it is almost
isotropic within a wide energy range%
\footnote{Namely, at $10~TeV < E_\nu < 3 \cdot 10^3~TeV$, the
maximum anisotropy is about 3--4\,\%~\cite{Bugaev89}.} and second, the
$\nu/\overline{\nu}$ ratio and the flavour ratio, $R$, are both about
1. These features provide a way to discriminate the PN contribution
through the analysis of the angular distribution and the relationship
between the muon and electron neutrino-induced events in a neutrino
telescope.

%%%%%%%%%%%%%%%%%%%%%%%%%%%%%%%%%%%%%%%%%%%%%%%%%%%%%%%%%%%%%%%%%%%%%%
%% Among the effects of this sort are the non-power-law spectrum and
%% energy-dependent chemical composition of the primary flux, growth
%% with energy of the total inelastic cross sections of hadron-nucleus
%% interactions, pion and kaon regeneration and overcharging,
%% production of nucleons, kaons and charmed hadrons in pion-nucleus
%% collisions, pion production in kaon decays, kaon and pion
%% production in charm decays, muon energy loss, muon polarization
%% (including the depolarization due to energy loss), meteorological
%% effects, etc. Below we will concern some of these points in short.
%%%%%%%%%%%%%%%%%%%%%%%%%%%%%%%%%%%%%%%%%%%%%%%%%%%%%%%%%%%%%%%%%%%%%

In this paper we calculate the conventional AN energy spectra with
taking into account one additional effect which was ignored in all
previous AN flux calculations, -- the dependence of the
$K^\pm_{\ell3}$ and $K^0_{\ell3}$ decay form factors on $q^2$ (where
$q$ is the 4-momentum transfer to the lepton pair). Inclusion of the
$q^2$-dependent form factors causes certain changes in the
differential $K_{\ell3}$ decay rates and therefore it should affect
the overall $\pi,K$ neutrino flux. Magnitude of the effect is
wittingly small, but it is energy-dependent and opposite in sign for
muon and electron neutrinos; consequently it must change predictions
for the $R$ as a function of energy. The $K_{\ell3}$ decay
contribution is ``by definition'' negligible at energies where the
prompt neutrinos become important. But, let us recall, the borderline
energy remains unknown for the present and hence the effect under
consideration might be interesting up to the PeV energy range.

In the next Section we present some necessary formulas from the theory
of neutrino production in the atmosphere. Section~\ref{sec:Kl3} deals
with the differential and total $K_{\ell3}$ decay rates. In
Section~\ref{sec:NC} we briefly describe our nuclear cascade model.
Section~\ref{sec:Res} is devoted to the discussion of our results for
the AN spectra, while the conclusions are in Section~\ref{sec:Conc}.
% Some auxiliary formulas are presented in Appendix.

\protect\section{Neutrino production in the atmosphere} \label{sec:GF}

Let $P$ labels a particle which can produce a lepton pair
$\overline{\ell}\nu_\ell$ or $\ell\overline{\nu}_\ell$ ($\ell = e,\mu$)
on decay and $\Phi_P\left(E_P,z,\vartheta\right)$ be the differential
energy spectrum of these particles as a function of energy $E_P$,
atmospheric depth $z$ and zenith angle $\vartheta$.
Let $d\Gamma_{P_{\ell k}}^\nu/d E_\nu$ be the differential with
respect to neutrino energy $E_\nu$ rate for the $k$-body decay mode
$P_{\ell k}$, as a function of $E_P$ and $E_\nu$ in the laboratory
frame of reference (the symbol $\nu$ stands for $\nu_\ell$ or
$\overline{\nu}_\ell$). Then the differential energy spectrum of
neutrinos at the depth $z$ and zenith angle $\vartheta$ is given by
\begin{equation} \label{1}
\Phi_\nu(E_\nu,z,\vartheta) =
\sum_P\sum_k\int_0^z\int_{E_{P_{\ell k}}}^\infty
\left[\frac{d\Gamma_{P_{\ell k}}^\nu\left(E_P,E_\nu\right)}
        {d E_\nu}\right]\Phi_P\left(E_P,z',\vartheta\right)
               \frac{d z'd E_P}{\rho\left(z',\vartheta\right)},
\end{equation}
where $\rho(z,\vartheta)$ is the air density on the corresponding
altitude in the atmosphere in terms of variables $z$ and
$\vartheta$; the summation is over all particles $P$ and $k$-body
decay modes.

The main decay modes answerable for the conventional neutrino
production are $\mu^\pm_{e3}$, $\pi^\pm_{\mu2}$, $K^\pm_{\mu2}$,
$K^\pm_{\ell3}$ and $K^0_{\ell3}$. Prompt neutrinos are produced
through the multiple modes of semileptonic decays of charmed hadrons
but, considering sizable gaps in the experimental data~\cite{PDG96}
and certain difficulties in the theoretical description of charm
decay, it is instructive (and generally accepted) to use the
inclusive approach. %to take all these modes into account.

The lower limit of integration over $E_P$ in eq.~(\ref{1}) is
defined by kinematics ($c = 1$). At $E_\nu \gg 1~GeV$,
\[
E_{P_{\ell k}} = \left(m_P^2/M^2_{P_{\ell k}}\right)E_\nu,
\]
where
\[
M^2_{\mu_{e  3}} = m_\mu^2,               \quad
M^2_{\pi_{\mu2}} = m_\pi^2-m_\mu^2,
\]
\[
M^2_{ K_{\mu 2}} = m_K^2-m_\mu^2,         \quad
M^2_{ K_{\ell3}} = m_K^2-m_\pi^2+m_\ell^2,
\]
and, for the inclusive decays,
\[
M^2_{P\rightarrow\overline{\ell}\nu X} = m_P^2+m_\ell^2-s_X^{\min}
\]
with $s_X^{\min}$ the invariant mass square minimum.

The total rate of the $P_{\ell k}$ decay in the lab{.} frame is
defined by
\[
\Gamma(P_{\ell k}) =
\int\left[\frac{d\Gamma_{P_{\ell k}}^\nu(E_P,E_\nu)}{d E_\nu}
 \right]d E_\nu = \frac{B\left(P_{\ell k}\right)m_P}{\tau_P E_P},
\]
where $m_P$ and $\tau_P$ are the mass and life time of the particle
$P$, respectively, and $B\left(P_{\ell k}\right)$ is the $P_{\ell k}$
decay branching ratio.  Let us introduce the ``spectral function''
\[
F_{P_{\ell k}}^\nu = \frac{E_P}{\Gamma\left({P_{\ell k}}\right)}
\left[\frac{d\Gamma_{P_{\ell k}}^\nu\left(E_P,E_\nu\right)}
                                               {d E_\nu}\right].
\]
It can be shown that, in the ultrarelativistic limit, the
$F_{P_{\ell k}}^\nu$ is a function of the only dimensionless
variable $x = E_\nu/E_P$ ($0 < x < M^2_{P_{\ell k}}/m_P^2$).

The spectral function for any two-body decay is merely constant.
In particular,
\[
F_{\pi_{\mu2}}^{\nu_\mu\left(\overline{\nu}_\mu\right)} =
                     \frac{1}{1-m_\mu^2/m_\pi^2}, \qquad
F_{  K_{\mu2}}^{\nu_\mu\left(\overline{\nu}_\mu\right)} =
                     \frac{1}{1-m_\mu^2/m_K  ^2}.
\]
The spectral functions for the three-particle decay of a polarized
muon in the ultrarelativistic limit are of the form~\cite{Naumov93}
\begin{eqnarray*}
F_{\mu_{e3}}^{\nu_e  \left(\overline{\nu}_e  \right)}(x) & = &
          2(1-x)^2\left[1+2x\pm{\cal P}_\mu(1-4x)\right], \\
F_{\mu_{e3}}^{\nu_\mu\left(\overline{\nu}_\mu\right)}(x) & = &
\frac{1}{3}(1-x)\left[5+5x-4x^2\pm{\cal P}_\mu(1+x-8x^2)\right],
\end{eqnarray*}
where ${\cal P}_\mu$ is the muon polarization dependent on the
muon and parent meson momenta. These dependencies are different for
the different meson decay modes. The $\mu$-decay contribution into
the $\nu_\mu+\overline{\nu}_\mu$ flux is very small at high energies
but in contrast, it dominates in the $\nu_e+\overline{\nu}_e$ flux up
to about 100~GeV for vertical and to several hundreds of GeVs for
horizontal directions. The muon polarization is therefore an
essential factor affecting the neutrino flavor ratio and the
neutrino to antineutrino ratio~\cite{Lipari93,Gaisser90}. However, at
$E_\nu > 1~TeV$ one can greatly simplify matter treating the
${\cal P}_\mu$ as an effective constant, $\langle{\cal P}_\mu\rangle$.
In our calculations we adopt $\langle{\cal P}_\mu\rangle = 0.33$.
Besides, as is customary in all AN flux calculations, we take no
account of a small change of the shape of neutrino distributions
which result from the radiative mode
$\mu\rightarrow e\overline{\nu}_e\nu_\mu\gamma$ (with branching
ratio of ($1.4 \pm 0.4)$\,\%) but simply increase the
$B\left(\mu_{e3}\right)$ to 100\,\%.

The spectral functions for the three-particle kaon decays calculated
without considering the $K_{\ell3}$ form factors can be found in
\cite{Volkova80} and \cite{Lipari93}. The inclusion of the
$q^2$-dependent form factors is the subject of the next Section.
One more point need to be made here. As in the case with $\mu$-decay,
below we neglect the radiative mode $K_L^0\rightarrow\pi
e\overline{\nu}_e\gamma$ (branching ratio is $(1.3\pm0.8)$\,\%) but
increase the $B\left(K_{e3}^0\right)$ to 40\,\% (cf.~\cite{PDG96}).
This approximation yields a completely negligible change in the
$\nu_e+\overline{\nu}_e$ flux.

In this paper, we will not enlarge on the calculation of the spectral
functions for the inclusive semileptonic decays of charmed hadrons.
Our simple phenomenological approach to the problem has been
described in \cite{Bugaev89}.

\protect\section{\protect{$K_{\ell3}$} decays} \label{sec:Kl3}

In the standard theory of weak interactions, the $K_{\ell3}$ decay
matrix element can be written in the form
\begin{equation} \label{2}
\frac{G_F}{\sqrt{2}}\sin\theta_C\left[f_+(q^2)(p_K+p_\pi)^\mu+f_-(q^2)
(p_K-p_\pi)^\mu\right]\overline{\ell}\gamma_\mu(1+\gamma_5)\nu_\ell.
\end{equation}
Here $G_F$ and $\theta_C$ are the Fermi constant and Cabibbo angle,
$p_{K,\pi}$ are the 4-momenta of the kaon and pion, and $f_\pm(q^2)$
are dimensionless form factors which are real functions of $q^2 =
(p_K-p_\pi)^2$, the square of the 4-momentum carried by leptons.
Experimental investigations of $K_{\ell3}$ decays suggest that
the form factors $f_\pm(q^2)$ are smooth functions of $q^2$ which
are normally written in the linear approximation as
\begin{equation}  \label{3}
f_\pm(q^2) = f_\pm(0)\left(1+\lambda_\pm\frac{q^2}{m_\pi^2}\right).
\end{equation}
In the limit of unbroken $SU(3)$ symmetry,
\begin{equation} \label{4}
f_+(0) = 1                  \qquad\text{for}\qquad K_{\ell3}^0
                            \qquad\text{and}\qquad
f_+(0) = \frac{1}{\sqrt{2}} \qquad\text{for}\qquad K_{\ell3}^\pm,
\end{equation}
while $f_-(0)$ reduces to zero. As a consequence, the parameter
$\xi = f_-(0)/f_+(0)$ should be small for $K_{e3}$ decays
(the Ademollo-Gatto theorem)~\cite{Kl3decay}. In the $K_{\mu3}$
case, the Ademollo-Gatto theorem is not valid and so, it is not
forbidden that $\xi\sim1$. The absolute normalization of the
$K_{\ell3}$ decay rates is not warranted for our purposes, as we use
the experimental values for $B(K_{\ell3})$ and $\tau_K$. This being
so, we adopt eqs.~(\ref{3},\ref{4}) from here on, considering
$\lambda_\pm$ and $\xi$ as input parameters.

From eqs.~(\ref{2}) and (\ref{3}), using standard
techniques~\cite{Brene61}, we find the differential (with respect to
neutrino energy) and total $K_{\ell3}$ decay rates in the lab{.} frame
of reference. In a general way, the $K_{\ell3}$ spectral function may
be written as~\cite{Naumov96}
\[
F_{K_{\ell3}}^\nu(x) = \frac{1}{Z}\sum_{n=-4}^3 C_n J_n(x),
\]
with the normalization constant
\[
Z = \sum_{n=-4}^3 C_n\left[J_{n+1}(0)-J_n(0)\right].
\]
Here
\[
J_n(x) = \int_{y_+}^{1-x}d y y^n\sqrt{(y-y_+)(y-y_-)}, \qquad
y_\pm  = \frac{(m_\pi \pm m_\ell)^2}{m_K^2},
\]
(these integrals are expressible in terms of elementary functions)
and $C_n$ are the constants proportional to $f_+^2(0)$ and dependent
on the masses, $m_K$, $m_\pi$, $m_\ell$ and the parameters
$\lambda_\pm$ and $\xi$. Let us define $C_n = f_+^2(0)c_n$.
Then the coefficients $c_n$ are
\begin{eqnarray*}
c_{-4}& = & -6r_\ell(r-r_\ell)^3v^2\lambda^2,                    \\
      &   &                                                      \\
c_{-3}& = & 8r_\ell(r-r_\ell)^2uv\lambda
                      -2(r-r_\ell)\{4(r-r_\ell)[r-r_\ell(1-4v)]  \\
      &   & -r_\ell[3r(1+3r)-r_\ell(9+10r-r_\ell)]v^2\}\lambda^2,\\
      &   &                                                      \\
c_{-2}& = & -3r_\ell(r-r_\ell)u^2+4\{2(r-r_\ell)[2(r-r_\ell)
                                                  +3r_\ell(u+v)] \\
      &   & -r_\ell[r+r_\ell+(r-r_\ell)(3+4r-r_\ell)]uv\}\lambda \\
      &   & +8\{(r-r_\ell)[(r-r_\ell)(4+3r+r_\ell)-(r+r_\ell)]
                      +4r_\ell[r(2+3r)-r_\ell(1+3r)]v\}\lambda^2 \\
      &   & -2r_\ell\{(r-r_\ell)[3(2+8r+3r^2)-r_\ell(2r+r_\ell)]
                         +(r+r_\ell)(3+3r+r_\ell)\}v^2\lambda^2, \\
      &   &                                                      \\
c_{-1}& = & -3\{4r-r_\ell[4(1-2u)+(1+r-r_\ell)u^2]\}
                                 -4\{8r(1+r)-4r_\ell(4+r+r_\ell) \\
      &   & +r_\ell[2(1+4r+r^2)-r_\ell(1+r+r_\ell)]uv
                                    -6r_\ell(1+2r)(u+v)\}\lambda \\
      &   & -8\{3r(1+3r+r^2)-r_\ell[5+16r+r^2+r_\ell(1+r+r_\ell) \\
      &   & +2(5+15r+6r^2+3r_\ell)v\}\lambda^2
                                      +2r_\ell\{3(1+r)(1+9r+r^2) \\
      &   & +r_\ell[3+4r+r^2+r_\ell(4+r+r_\ell)]\}v^2\lambda^2,  \\
      &   &                                                      \\
c_0   & = & 3\{4(1+r)-r_\ell[4+(u-8)u]\}+4\{4(1+4r+r^2)          \\
      &   & -r_\ell[4(5-r+2r_\ell)-6(2+r+r_\ell)(u+v)
                                      +(4+4r+r_\ell)uv]\}\lambda \\
      &   & +8\{(1+r)(1+8r+r^2-r_\ell[3-4r-r^2+r_\ell
                                       (3-r+3r_\ell)]\}\lambda^2 \\
      &   & +2r_\ell\{16[3+6r+r^2+r_\ell(3+r+r_\ell)]
                 -[9(1+3r+r^2)+r_\ell(7+4r+r_\ell)]v\}v\lambda^2,\\
      &   &                                                      \\
c_1   & = & -12-8\{4(1+r)-r_\ell[2-3(u+v)-uv]\}\lambda
                                                -2\{12(1+3r+r^2) \\
      &   & -r_\ell[4(7+5r_\ell)-32(3+2r+2r_\ell)v
                                  +(9+9r+5r_\ell)v^2]\}\lambda^2,\\
      &   &                                                      \\
c_2   & = & 16\lambda+2\{12(1+r)-r_\ell[4-(16-3v)v]\}\lambda^2,  \\
      &   &                                                      \\
c_3   & = & -8\lambda^2,
\end{eqnarray*}
where we used the following notation:
\[
r = \frac{m_\pi^2}{m_K^2},            \qquad
r_\ell = \frac{m_\ell^2}{m_K^2},
\]
and
\[
u = 1-\xi,                            \qquad
v = 1-\xi\frac{\lambda_-}{\lambda_+}, \qquad
\lambda = \frac{\lambda_+}{2r}.
\]
To estimate the numerical values for the $C_n$, we use the
parameters $\lambda_+$ and $\xi$ evaluated by the Particle Data
Group~\cite{PDG96} (see table~\ref{t:4}).
%--------------------------------------------------------------------
\begin{table}[htb]
\protect\caption{$K_{\ell 3}$ form factor parameters \protect\cite{PDG96}.}
\label{t:4}
\begin{tabular}{lcccc}
            & $K^0  _{e3}       $ & $K^0  _{\mu3}   $
            & $K^\pm_{e3}       $ & $K^\pm_{\mu3}   $ \\ \hline
$\lambda_+$ & $0.0300 \pm 0.0016$ & $0.034 \pm 0.005$
            & $0.0286 \pm 0.0022$ & $0.033 \pm 0.008$ \\
$\xi$       & \multicolumn{2}{c}{$-0.11 \pm 0.09$}
            & \multicolumn{2}{c}{$-0.35 \pm 0.15$}    \\
\end{tabular}
\end{table}
%--------------------------------------------------------------------

Since the term of the matrix element~(\ref{2}) proportional to $f_-$
can be neglected for $K_{e3}$ and most $K_{\mu3}$ data are adequately
described with a constant $f_-$~\cite{PDG96}, in the subsequent
discussion we assume $\lambda_- = 0$. The numerical values of the
coefficients $C_n$ are given in tables~\ref{t:5} and~\ref{t:6}. For
comparison we also included in these tables the $C_n$ calculated with
$\lambda_+ = 0$.

In the $K_{e3}$ case, electron mass can be neglected as compared to
pion mass. In this approximation, one can easily found
\begin{eqnarray} \label{5a}
F_{K_{e3}}^{\nu_e}(x)
          & = & \frac{1}{Z}\left\{\frac{rC_{-3}}{2(1-x)^2}+
                \frac{rC_{-2}-C_{-3}}{1-x}+(C_{-2}-rC_{-1})
                          \ln(1-x)+C_{-1}(1-x)\right. \nonumber\\
          &   & +\sum_{n=0}^3 C_n\left[(1-x)^{n+1}
                 \left(\frac{1-x}{n+2}-\frac{r}{n+1}\right)
                +\frac{r^{n+2}}{(n+1)(n+2)}\right]    \nonumber\\
          &   & \left.+\frac{C_{-3}}{2r}-C_{-2}(\ln r+1)
                +rC_{-1}(\ln r-1)\right\},
\end{eqnarray}

%--------------------------------------------------------------------
\begin{table}[htb]
\caption{Coefficients \protect{$C_n$} for \protect{$K_{\ell3}^0$}
         decays (\protect{$\lambda_-=0$}, \protect{$\xi=-0.11$}).}
\label{t:5}
\begin{tabular}{ccccc}
         & \multicolumn{2}{c}{$K_L^0\rightarrow\pi^\pm
          e^\mp\overline{\nu}_e  \left(\nu_e  \right)$}
         & \multicolumn{2}{c}{$K_L^0\rightarrow\pi^\pm
        \mu^\mp\overline{\nu}_\mu\left(\nu_\mu\right)$}\\\cline{2-5}
$C_n   $ & $\lambda_+=0$ & $\lambda_+=0.030$
         & $\lambda_+=0$ & $\lambda_+=0.034$           \\\hline
$C_{-4}$ & $  0               $ & $  0               $
         & $  0               $ & $ -1.99\cdot10^{-8}$ \\
$C_{-3}$ & $  0               $ & $ -5.90\cdot10^{-6}$
         & $  0               $ & $  3.78\cdot10^{-6}$ \\
$C_{-2}$ & $  0               $ & $  1.03\cdot10^{-3}$
         & $ -2.29\cdot10^{-4}$ & $  4.33\cdot10^{-4}$ \\
$C_{-1}$ & $ -3.93\cdot10^{-2}$ & $ -6.45\cdot10^{-2}$
         & $ -5.93\cdot10^{-2}$ & $ -7.94\cdot10^{-2}$ \\
$C_{ 0}$ & $  5.39\cdot10^{-1}$ & $  7.29\cdot10^{-1}$
         & $  5.60\cdot10^{-1}$ & $  7.80\cdot10^{-1}$ \\
$C_{ 1}$ & $ -5.00\cdot10^{-1}$ & $ -8.19\cdot10^{-1}$
         & $ -5.00\cdot10^{-1}$ & $ -8.82\cdot10^{-1}$ \\
$C_{ 2}$ & $  0               $ & $  1.66\cdot10^{-1}$
         & $  0               $ & $  1.96\cdot10^{-1}$ \\
$C_{ 3}$ & $  0               $ & $ -1.21\cdot10^{-2}$
         & $  0               $ & $ -1.56\cdot10^{-2}$ \\
\end{tabular}
\vspace*{0.4cm}
%\end{table}
%\begin{table}
\caption{Coefficients \protect{$C_n$} for \protect{$K_{\ell3}^\pm$}
         decays (\protect{$\lambda_-=0$}, \protect{$\xi=-0.35$}).}
\label{t:6}
\begin{tabular}{ccccc}
         & \multicolumn{2}{c}{$K^\pm\rightarrow\pi^0  e^\pm\nu_e
                      \left(\overline{\nu}_e  \right)$}
         & \multicolumn{2}{c}{$K^\pm\rightarrow\pi^0\mu^\pm\nu_\mu
                      \left(\overline{\nu}_\mu\right)$}\\\cline{2-5}
$C_n   $ & $\lambda_+=0$ & $\lambda_+=0.0286$
         & $\lambda_+=0$ & $\lambda_+=0.0330$          \\\hline
$C_{-4}$ & $  0               $ & $  0               $
         & $  0               $ & $ -6.77\cdot10^{-9}$ \\
$C_{-3}$ & $  0               $ & $ -2.55\cdot10^{-6}$
         & $  0               $ & $  1.95\cdot10^{-6}$ \\
$C_{-2}$ & $  0               $ & $  4.66\cdot10^{-4}$
         & $ -1.51\cdot10^{-4}$ & $  1.17\cdot10^{-4}$ \\
$C_{-1}$ & $ -1.87\cdot10^{-2}$ & $ -3.06\cdot10^{-2}$
         & $ -3.28\cdot10^{-2}$ & $ -4.31\cdot10^{-2}$ \\
$C_{ 0}$ & $  2.69\cdot10^{-1}$ & $  3.62\cdot10^{-1}$
         & $  2.83\cdot10^{-1}$ & $  3.96\cdot10^{-1}$ \\
$C_{ 1}$ & $ -2.50\cdot10^{-1}$ & $ -4.10\cdot10^{-1}$
         & $ -2.50\cdot10^{-1}$ & $ -4.46\cdot10^{-1}$ \\
$C_{ 2}$ & $  0               $ & $  8.34\cdot10^{-2}$
         & $  0               $ & $  1.01\cdot10^{-1}$ \\
$C_{ 3}$ & $  0               $ & $ -6.10\cdot10^{-3}$
         & $  0               $ & $ -8.12\cdot10^{-3}$ \\
\end{tabular}
\end{table}
%--------------------------------------------------------------------
\noindent
where
\begin{eqnarray} \label{5b}
Z & = & \left[(1+r)C_{-2}-C_{-3}-rC_{-1}\right]\ln r+2(1-r)C_{-2}
       -\frac{1-r^2}{2r}\left(C_{-3}+rC_{-1}\right)\nonumber\\
  &   & +\sum_{n=0}^3 C_n\left[\frac{r(1-r^{n+1})}{(n+1)(n+2)}
        -\frac{1-r^{n+3}}{(n+2)(n+3)}\right]
\end{eqnarray}
and $r = {m_\pi^2}/{m_K^2}$. The total $K_{e3}$ decay rate in the
rest frame has the form (cf.~\cite{Kl3decay})
\begin{equation} \label{6}
\Gamma^*(K_{e3}) = \frac{G_F^2\sin^2\theta_C m_K^5}{768\pi^3}
                f_+^2(0)\left(a_0+a_1\lambda_++a_2\lambda_+^2\right),
\end{equation}
with constant $a_i$. Substituting the numerical values of the
parameters yields
\begin{eqnarray*}
a_0 \simeq 0.576, \quad a_1 \simeq 2.140, \quad
a_2 \simeq 1.580  & \quad\text{for}\quad & K^\pm_{e3}, \\
a_0 \simeq 0.560, \quad a_1 \simeq 1.947, \quad
a_2 \simeq 1.345  & \quad\text{for}\quad & K^0_{e3}.
\end{eqnarray*}

\begin{figure}[htb]
 \center\mbox{\epsfig{file=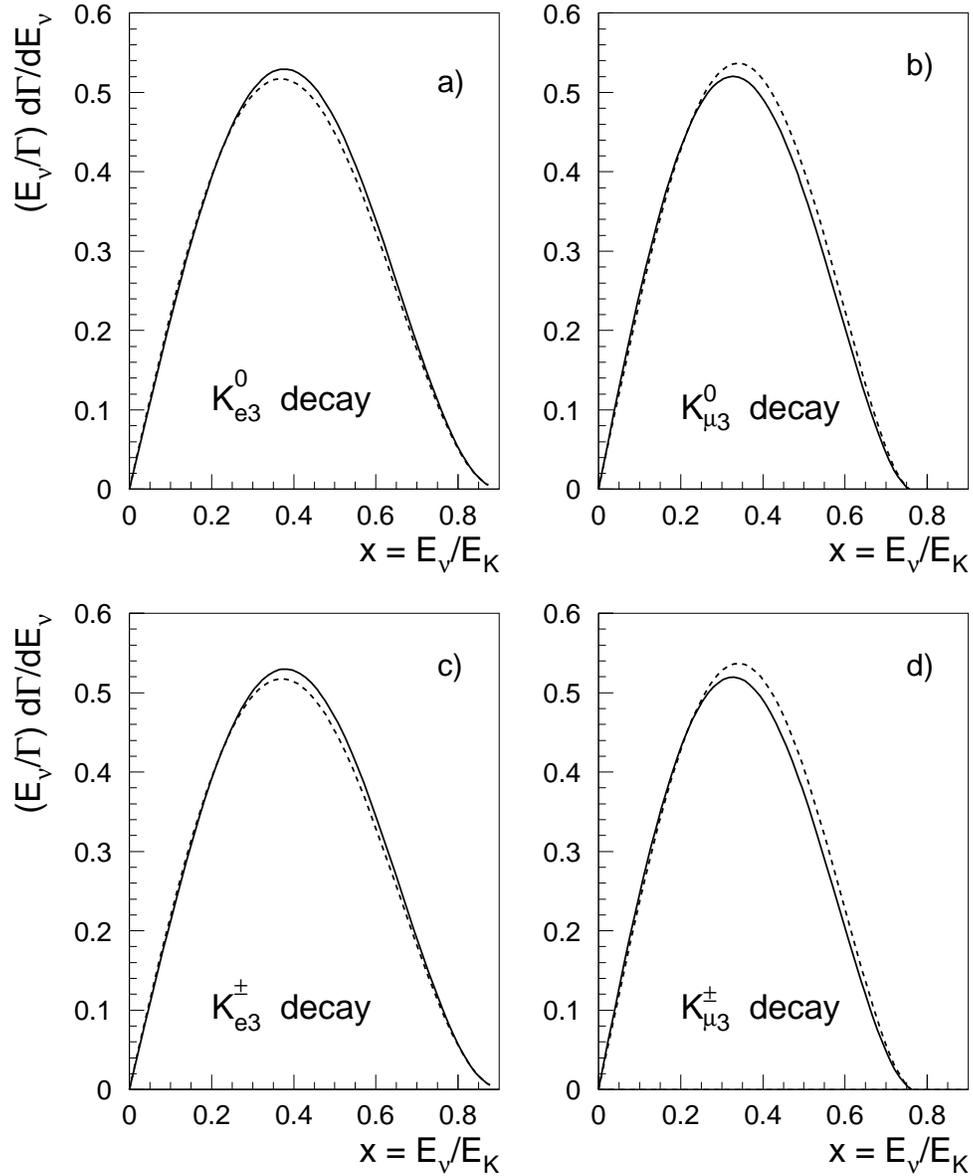,height=16.5cm}}
 \protect\caption{The normalized $x$-distributions of
                 (anti)neutrinos in $K_{\ell 3}$ decays,
                 $xF_{K_{\ell3}}^\nu(x)$, calculated with
                 $f_+ = f_+(0)$ (dashed curves) and with
                 $q^2$-dependent $f_+$ (solid curves).
 \label{f:1}}
 \end{figure}

\begin{figure}[htb]
 \center\mbox{\epsfig{file=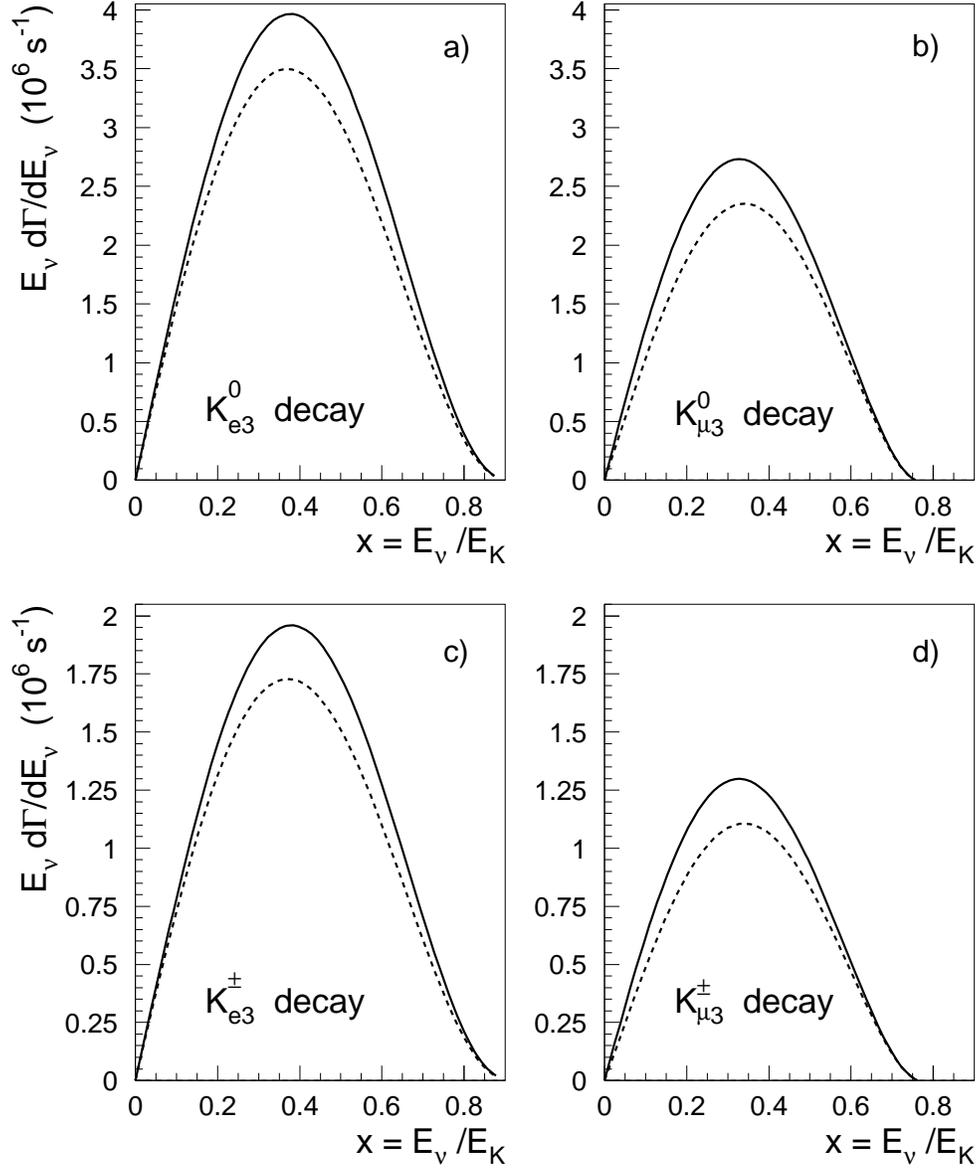,height=16.5cm}}
 \protect\caption{The absolute $x$-distributions of
                 (anti)neutrinos in $K_{\ell 3}$ decays,
                 $E_\nu d\Gamma_{K_{\ell k}}^\nu/d E_\nu$,
                 calculated with $f_+ = f_+(0)$ (dashed curves) and
                 with $q^2$-dependent $f_+$ (solid curves).
 \label{f:2}}
 \end{figure}

Let us pass over the explicit form of the $F_{K_{\mu3}}^{\nu_\mu}(x)$
and $\Gamma^*(K_{\mu3})$ which is much more complicated in comparison
with eqs.~(\ref{5a}--\ref{6}), and proceed to some numerical
results.  Figures~\ref{f:1} and~\ref{f:2} show respectively the
$K_{\ell3}$ spectral functions and differential decay rates (versus
$x$) calculated with $f_+ = f_+(0)$ and with the $q^2$-dependent form
factors. As is seen from fig.~\ref{f:1}, the effect for the spectral
functions is different in magnitude and, more importantly, opposite
for electron and muon neutrinos, even though the absolute
differential rates (fig.~\ref{f:2}) grow if the form factors are
accounted for. As a result, the $K_{\ell3}$ contribution to the AN
flux shall slightly be increased for electron neutrinos and decreased
for muon neutrinos, compared to the case of constant form factors.
Clearly, in the range where the conventional neutrinos dominate, the
magnitude of the effect in the overall AN flux is determined by the
kaon production cross sections that is by the ``$K/\pi$ ratio'' and,
to a lesser extend, by the cross sections for kaon regeneration.

In table~\ref{t:7} we give the decay rates, $\Gamma^*(K_{\ell3})$,
calculated using the above formulas with $f_+ = f_+(0)$ and with the
linear dependence of $f_+$ on $q^2$~(\ref{3}), together with the best
fits of experimental data obtained by the Particle Data
Group~\cite{PDG96}.
%--------------------------------------------------------------------
\begin{table}[htb]
\caption{\protect{$K_{\ell 3}$} decay rates.}
\label{t:7}
\begin{tabular}{cccc}
Decay & Calculated with & Calculated with & Experimental \\
mode  & $f_+=f_+( 0 )$ ($10^6$ s$^{-1}$)
      & $f_+=f_+(q^2)$ ($10^6$ s$^{-1}$)
      & best fit~\protect\cite{PDG96} ($10^6$ s$^{-1}$)  \\ \hline
$K^0  _{e  3}$ & 6.76  & 7.49 & $7.49 \pm 0.06$          \\
$K^0  _{\mu3}$ & 4.38  & 5.25 & $5.25 \pm 0.05$          \\
$K^\pm_{e  3}$ & 3.34  & 3.70 & $3.89 \pm 0.05$          \\
$K^\pm_{\mu3}$ & 2.06  & 2.50 & $2.57 \pm 0.06$          \\
\end{tabular}
\end{table}
%--------------------------------------------------------------------

As table~\ref{t:7} suggests, the inclusion of the $q^2$-dependent form
factors causes the increase of about 11\,\% in the $K_{e3}$ decay
rates and of about 20\,\% in the $K_{\mu3}$ rates. The improved rates
correlate well with the experimental data except the $K^\pm_{e3}$
case. But the latter is not an essential flaw, having regard to the
variance of the world data on $\Gamma^*(K^\pm_{e3})$ which is far more
than the error of the PDG best fit. The additional corrections due to
the $SU(3)$ symmetry breaking should be less than 3--4\,\% and, what
is more essential for our study, they cannot change the spectral
functions, $F_{K_{\ell3}}^\nu(x)$, that is the shape of the neutrino
distributions from $K_{\ell3}$ decays.

\protect\section{Nuclear cascade model} \label{sec:NC}

Our nuclear cascade calculations are based on the model by Vall
et al.~\cite{Vall86}. The results obtained within this model
agree well with all available experimental data on hadron spectra
(including the single proton, neutron and pion fluxes) for various
atmospheric depths at energies from 1~TeV up to about 600~TeV.
The muon spectrum calculated with this model is in reasonable
agreement with the current sea-level and underground
data~\cite{Bugaev93}. It is also in very good agreement
(within 5\,\%) with the recent Monte Carlo calculation by Agrawal
et al.~\cite{Agrawal96}.

The model takes into account the processes of regeneration and
recharging of nucleons and pions as well as production of kaons,
nucleons and charmed hadrons ($D^\pm$, $D^0$, $\overline{D}{}^0$,
$\Lambda_c^+$) in pion--nucleus collisions, muon energy loss, etc.
Let us briefly enumerate here the basic assumptions and discuss some
pluses and minuses of the model. Further details and numerical values
of the input parameters can be found in \cite{Vall86} for $\pi$ and
$K$ meson production and in \cite{Bugaev89} for charm production.

(\emph{i}) Our (analytical) model of the all-particle primary spectra
and chemical composition is taken from Nikolsky et al.~\cite{Nikolsky84}.
This model adequately describes the world data on primary cosmic rays
from about 1~TeV/nucleon up to the range well beyond the ``knee''
region (see a discussion in Bugaev et al.~\cite{Bugaev93}). The nuclear
component of the primary flux is treated on the basis of the
superposition model.

(\emph{ii}) We assume a logarithmic growth with energy of the total
inelastic cross sections $\sigma_{hA}^{\it{inel}}$ for
interactions of a hadron $h$ with a nuclear target $A$. Such
dependence arises from a model for elastic amplitude of
hadron--hadron collisions, based on the conception of double pomeron
with supercritical intercept. For simplicity sake we also use another
consequence of this model, the asymptotic equality of the inelastic
cross sections for any projectile hadron. Namely, we assume
\[
\sigma_{hA}^{\it{inel}}(E_h) = \sigma_{hA}+\sigma_A\ln(E_h/E_1),
              \qquad\text{for}\quad h = N, \pi, K, D, \Lambda_c,
\]
($A$ is the ``air nucleus'') at $E_h > E_1 = 1~TeV$. The calculated
cross sections for $h = N, \pi, K$ are in reasonable agreement with
available accelerator and cosmic ray data. There are no the data on
the inelastic cross sections for charmed hadrons but we notice that
the prompt lepton flux is scarcely affected by the specific values of
these cross sections up to about $10^4~TeV$ of lepton energy (due to
very short lifetime of charmed particles). Thus even a rough
estimation of $\sigma_{D A}^{\it{inel}}$ and
$\sigma_{\Lambda_c A}^{\it{inel}}$ will suffice for our purposes.

(\emph{iii}) We assume Feynman scaling in the fragmentation range for
the inclusive processes $h_i A \rightarrow h_f X$, with $h_i = p, n,
\pi^\pm$ and $h_f = p, n, \pi^\pm, K^\pm, K^0, \overline{K}{}^0$.
Thus the invariant inclusive cross sections
$Ed^3\sigma_{fi}^A/d^3p$ are energy independent at large Feynman
$x$. The truth of this assumption is an outstanding question.

(\emph{iv}) The kaon regeneration (i.e.~  the processes
$K^\pm A \rightarrow K^\pm X$, $K^\pm A \rightarrow K^0 X$, and so
on) is disregarded in our calculations. We also neglect nucleon, pion
and charm production in kaon--nucleus collisions as well as pion
production in kaon decays, which makes it possible to study the
``$\pi N$'' cascade without reference to kaons. All these
simplifications yield a somewhat conservative result for the
conventional AN spectra and must be avoided in the future study;
the proper inclusion of the kaon regeneration is the most essential
point.

(\emph{v}) At the stage of nuclear cascade (but of course not at the
lepton production stage) the decay of charged pions is neglected.
This approximation greatly simplifies the description of the pion
regeneration/overcharging and nucleon production in pion--nucleus
collisions and it is valid at the pion energies above 1--2~TeV
for directions close to vertical. However it becomes too crude for
near-horizontal directions at $E_\nu < 7-8~TeV$. To extend our
results up to 1~TeV, we included the appropriate corrections for
pion decay using a numerical procedure.

\protect\section{AN flux (numerical results)} \label{sec:Res}

\begin{figure}[htb]
 \center\mbox{\epsfig{file=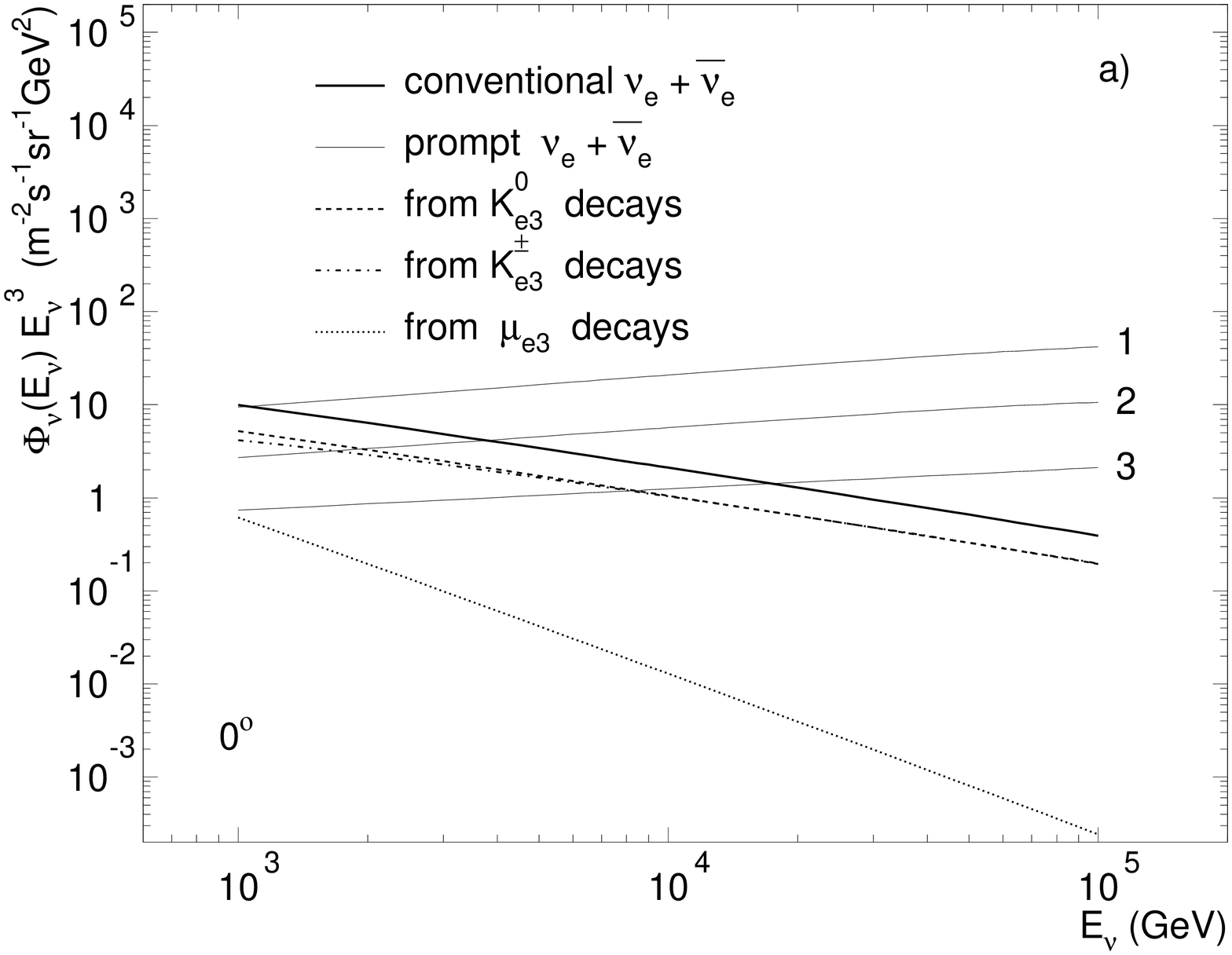,height=9.1cm}}
 \center\mbox{\epsfig{file=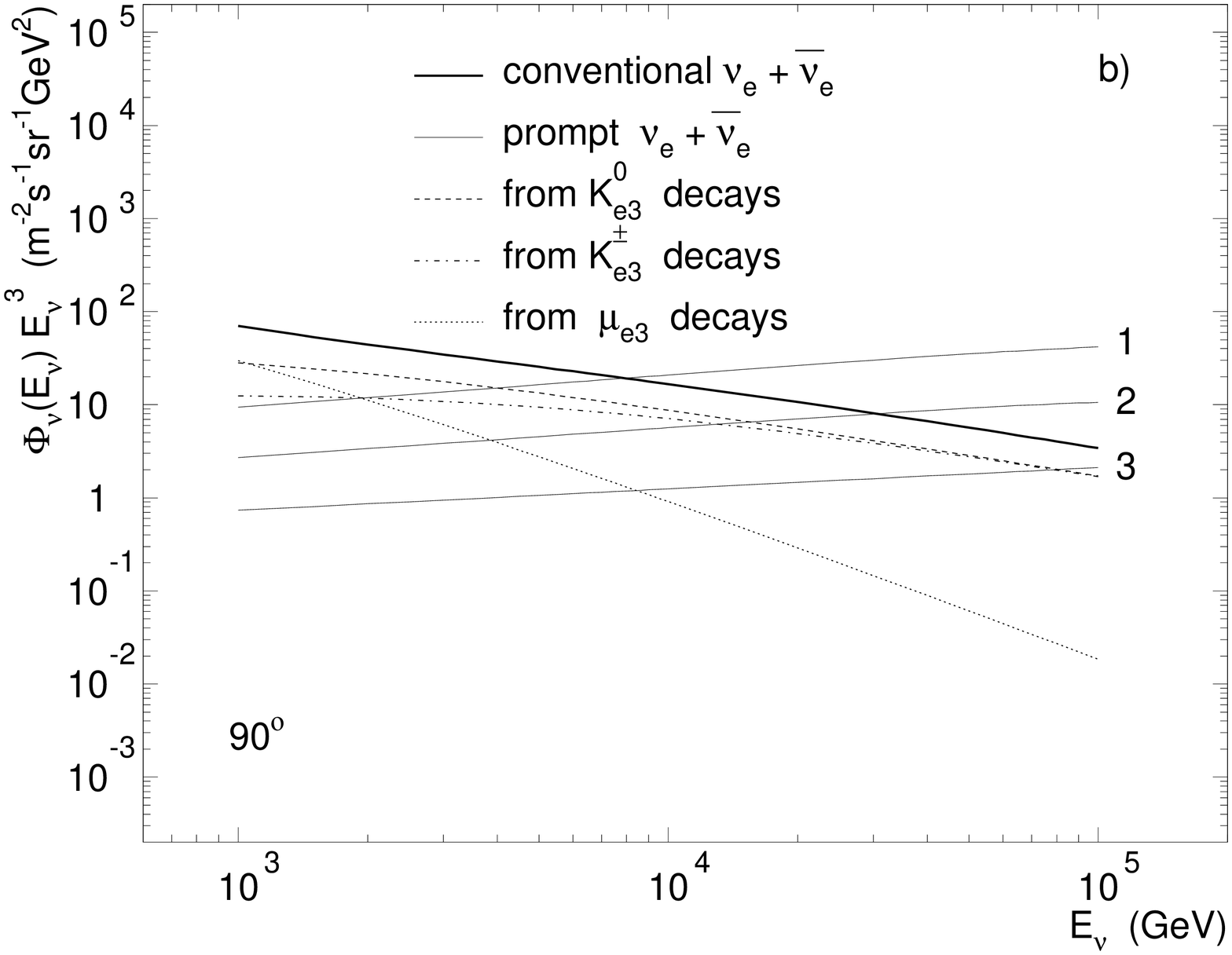,height=9.1cm}}
 \protect\caption{Different contributions to the vertical (a) and
                 horizontal (b) electron neutrino + antineutrino
                 fluxes. The PN contributions calculated in RQPM,
                 QGSM and pQCD are marked ``1'', ``2'' and ``3'',
                 respectively.
 \label{f:3}}
 \end{figure}

\begin{figure}[htb]
 \center\mbox{\epsfig{file=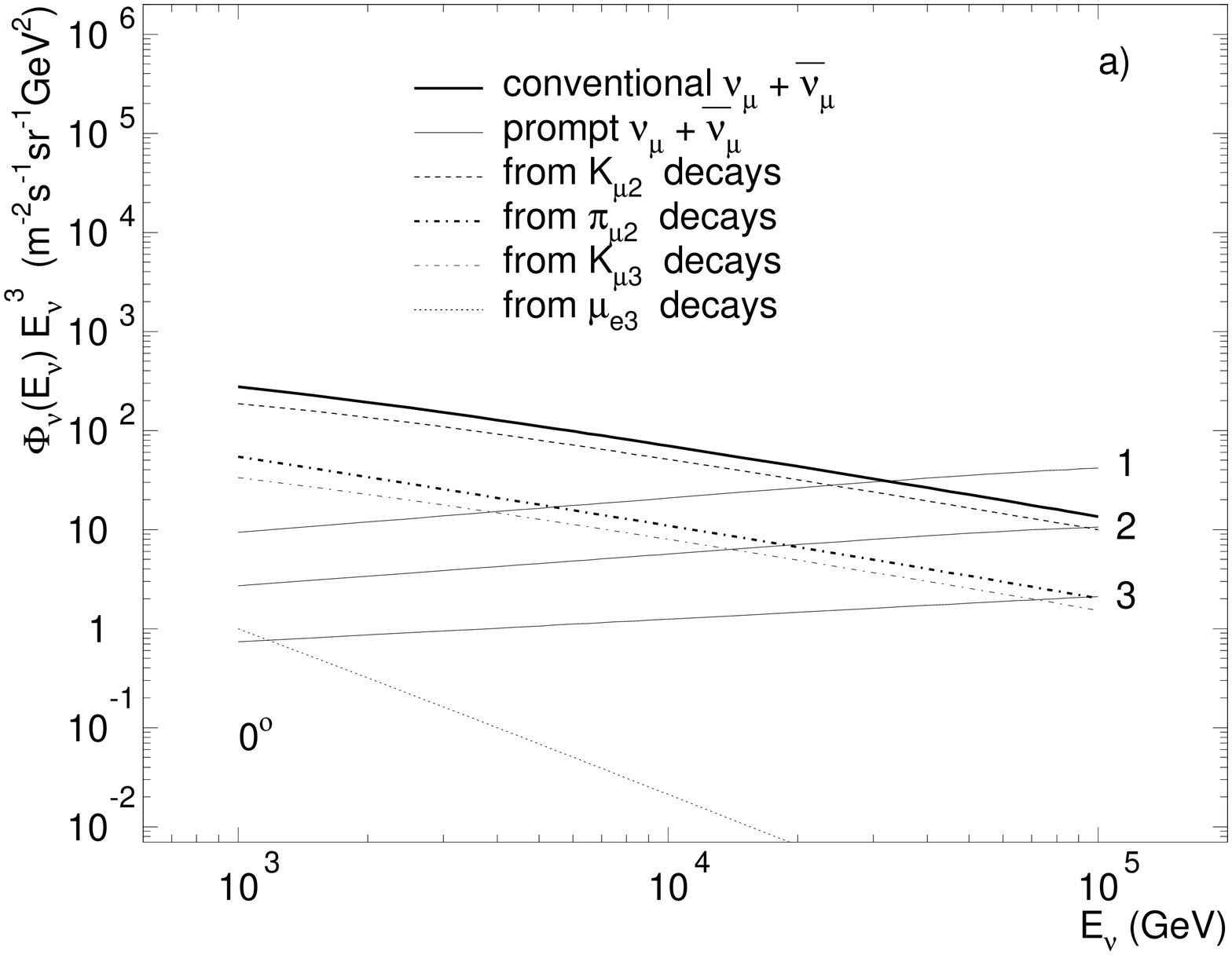,height=9.1cm}}
 \center\mbox{\epsfig{file=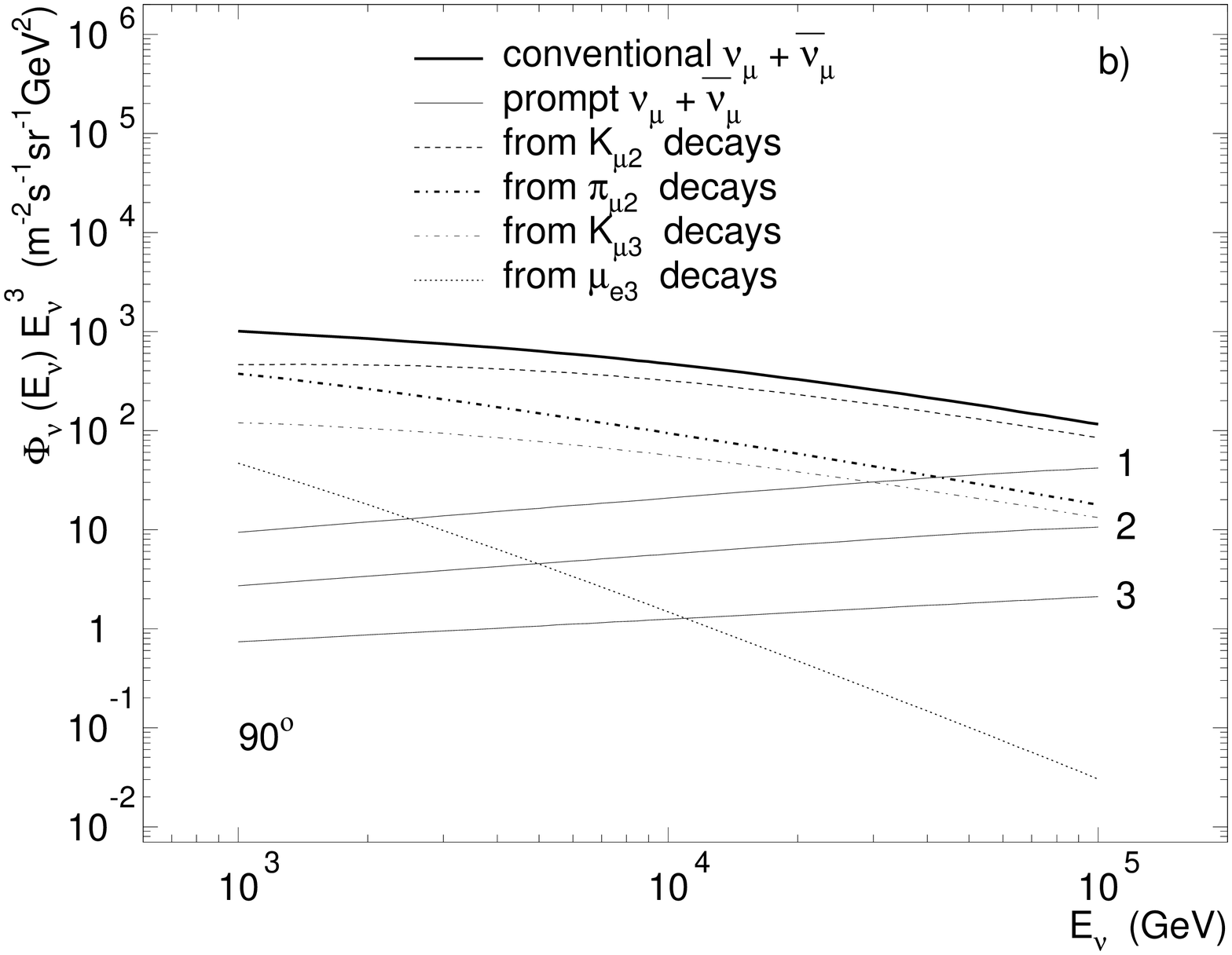,height=9.1cm}}
 \protect\caption{Different contributions to the vertical (a) and
                 horizontal (b) muon neutrino + antineutrino fluxes.
                 The PN contributions calculated in RQPM, QGSM and
                 pQCD are marked ``1'', ``2'' and ``3'', respectively.
 \label{f:4}}
 \end{figure}

\begin{figure}[htb]
 \center\mbox{\epsfig{file=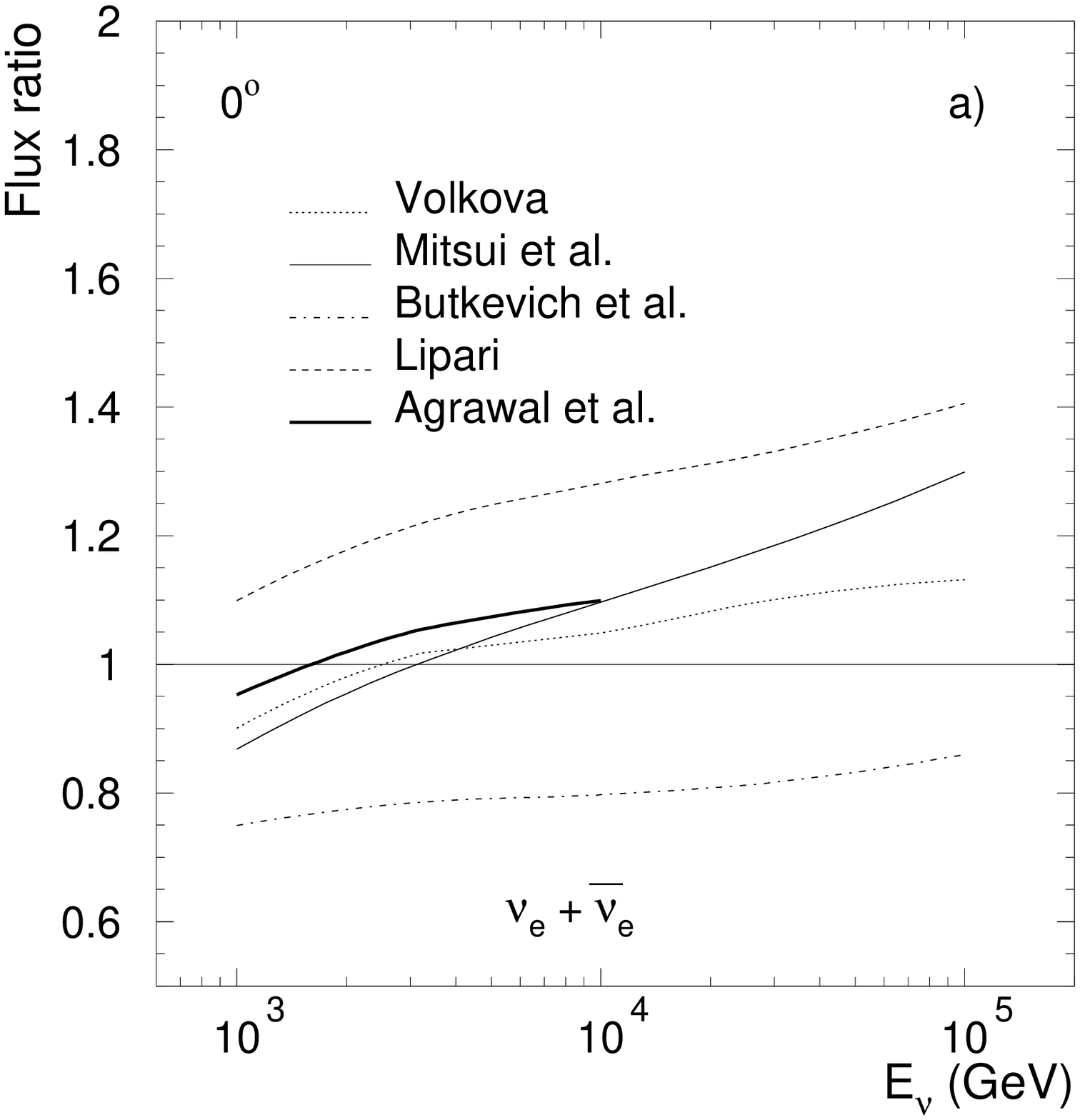,width=6.4cm,height=6.4cm}
     \hspace{2mm}\epsfig{file=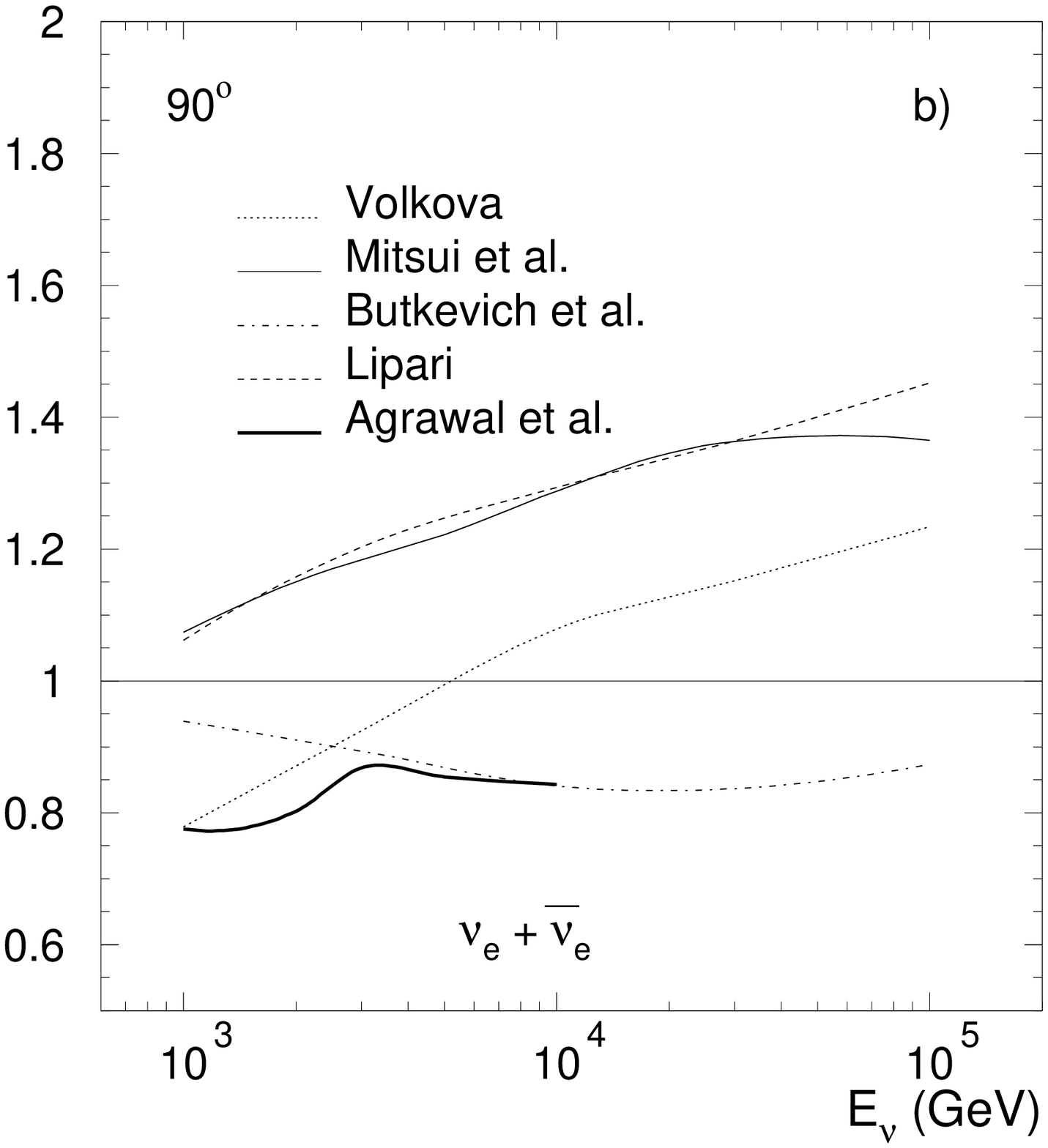,width=6.4cm,height=6.4cm}}
 \center\mbox{\epsfig{file=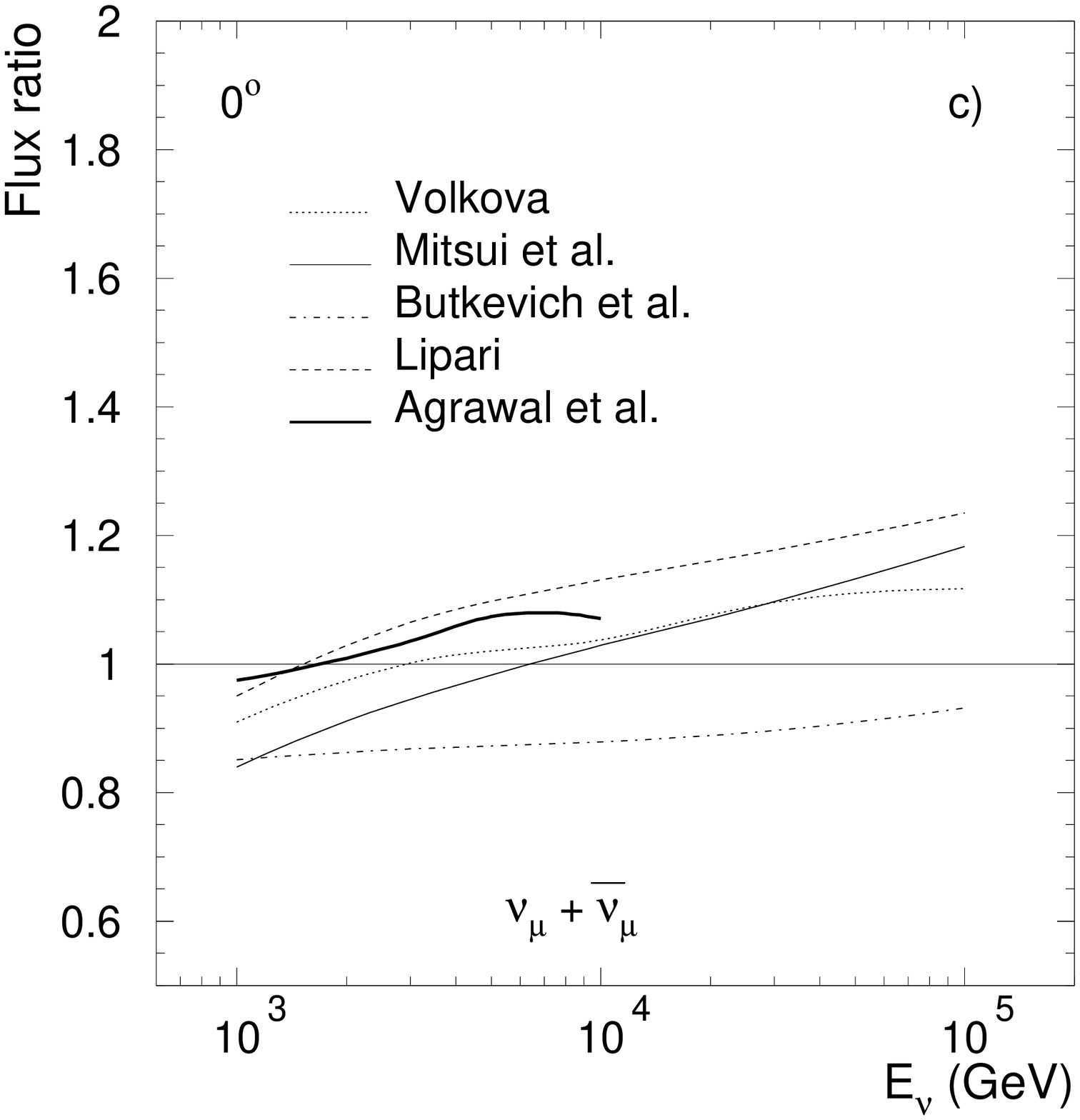,width=6.4cm,height=6.4cm}
     \hspace{2mm}\epsfig{file=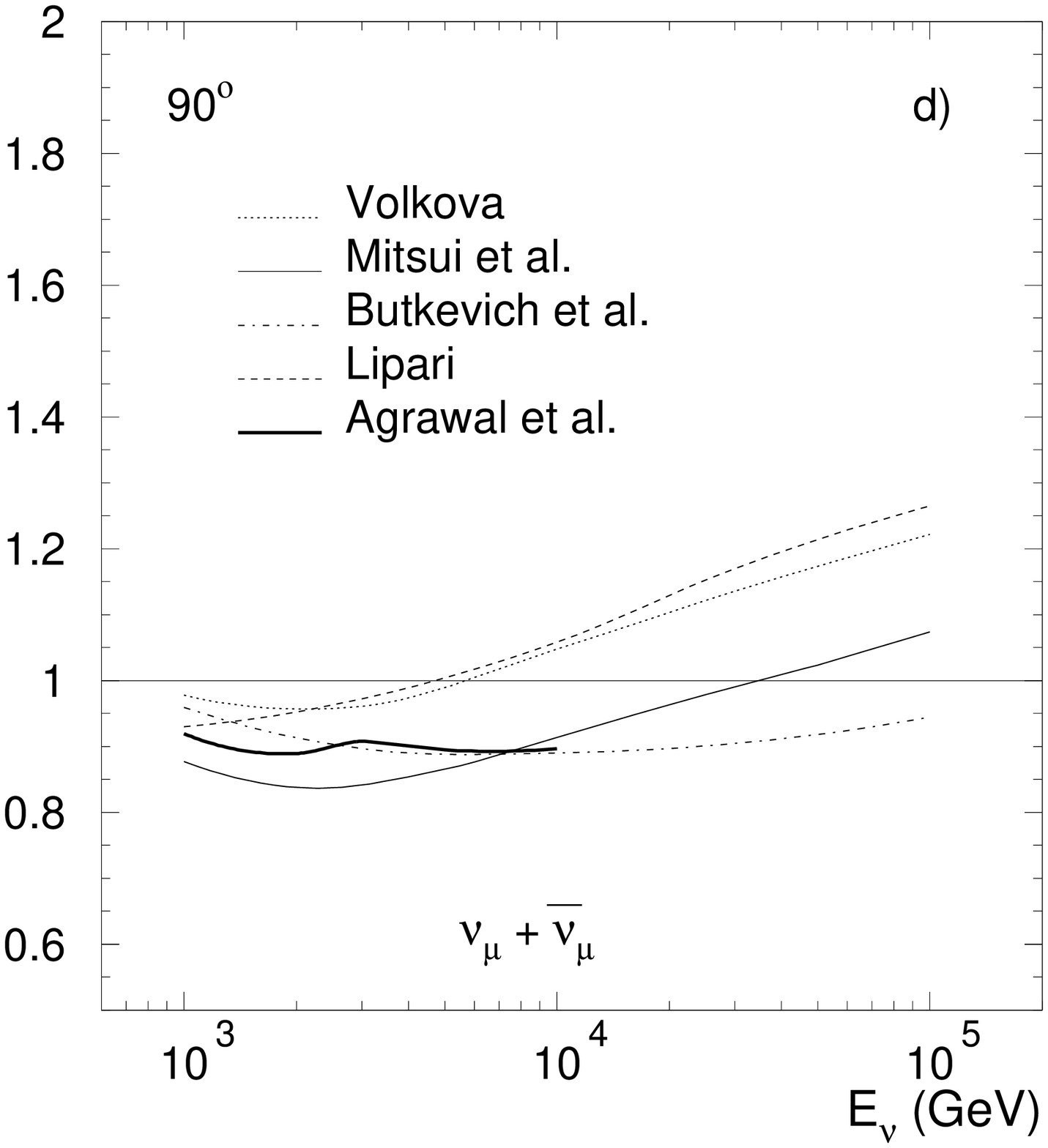,width=6.4cm,height=6.4cm}}
 \protect\caption{Conventional $\nu_e+\overline{\nu}_e$ and
                 $\nu_\mu+\overline{\nu}_\mu$ fluxes at
                 $\vartheta = 0^\circ$ and $90^\circ$ from
                 Volkova~\protect\cite{Volkova80},
                 Mitsui et al.~\protect\cite{Mitsui86},
                 Butkevich et al.~\protect\cite{Butkevich89},
                 Lipari~\protect\cite{Lipari93} and
                 Agrawal et al.~\protect\cite{Agrawal96},
                 normalized to the fluxes calculated in this work.
 \label{f:5}}
 \end{figure}

\begin{figure}[htb]
 \center\mbox{\epsfig{file=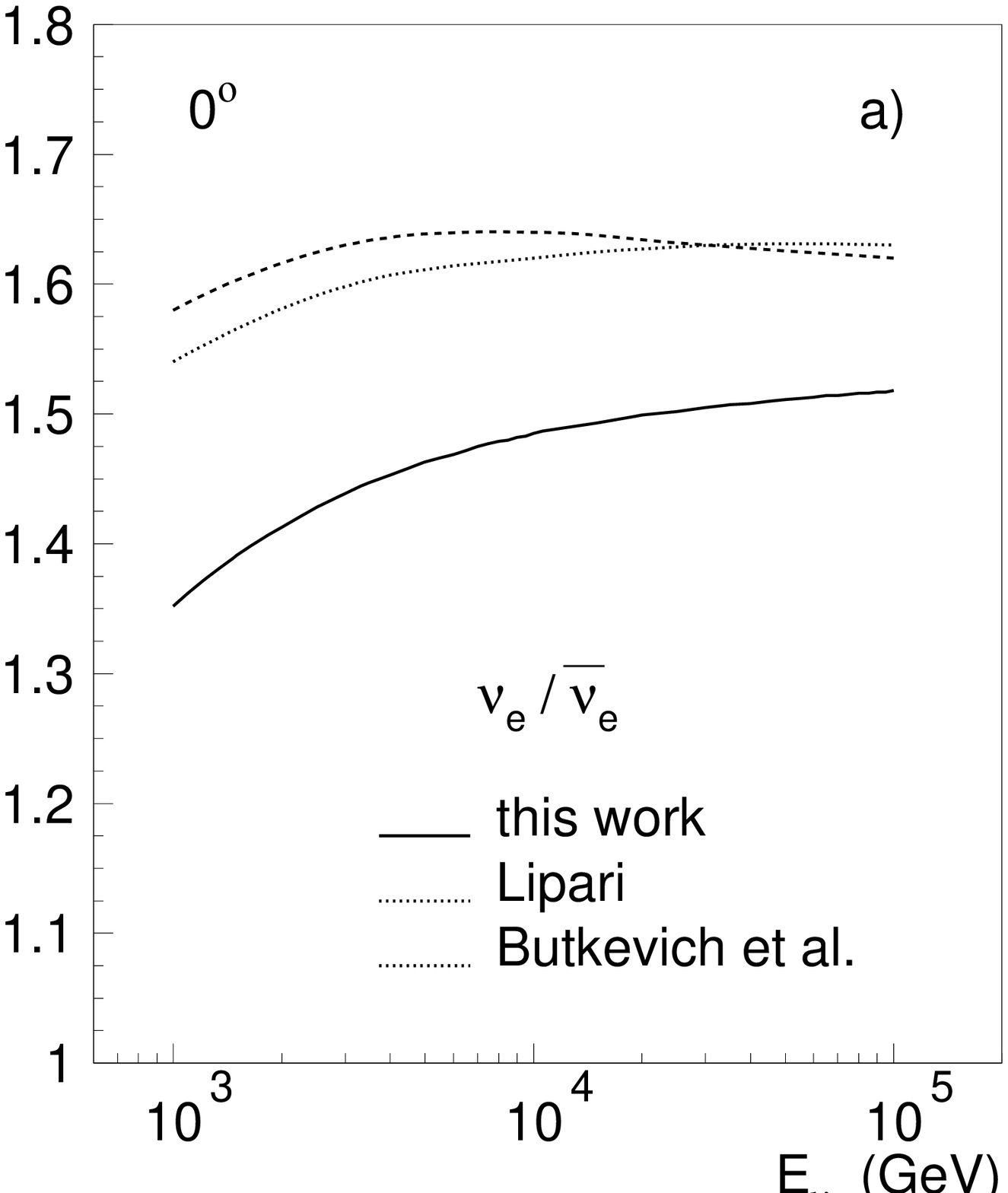,width=6.4cm,height=6.4cm}
     \hspace{2mm}\epsfig{file=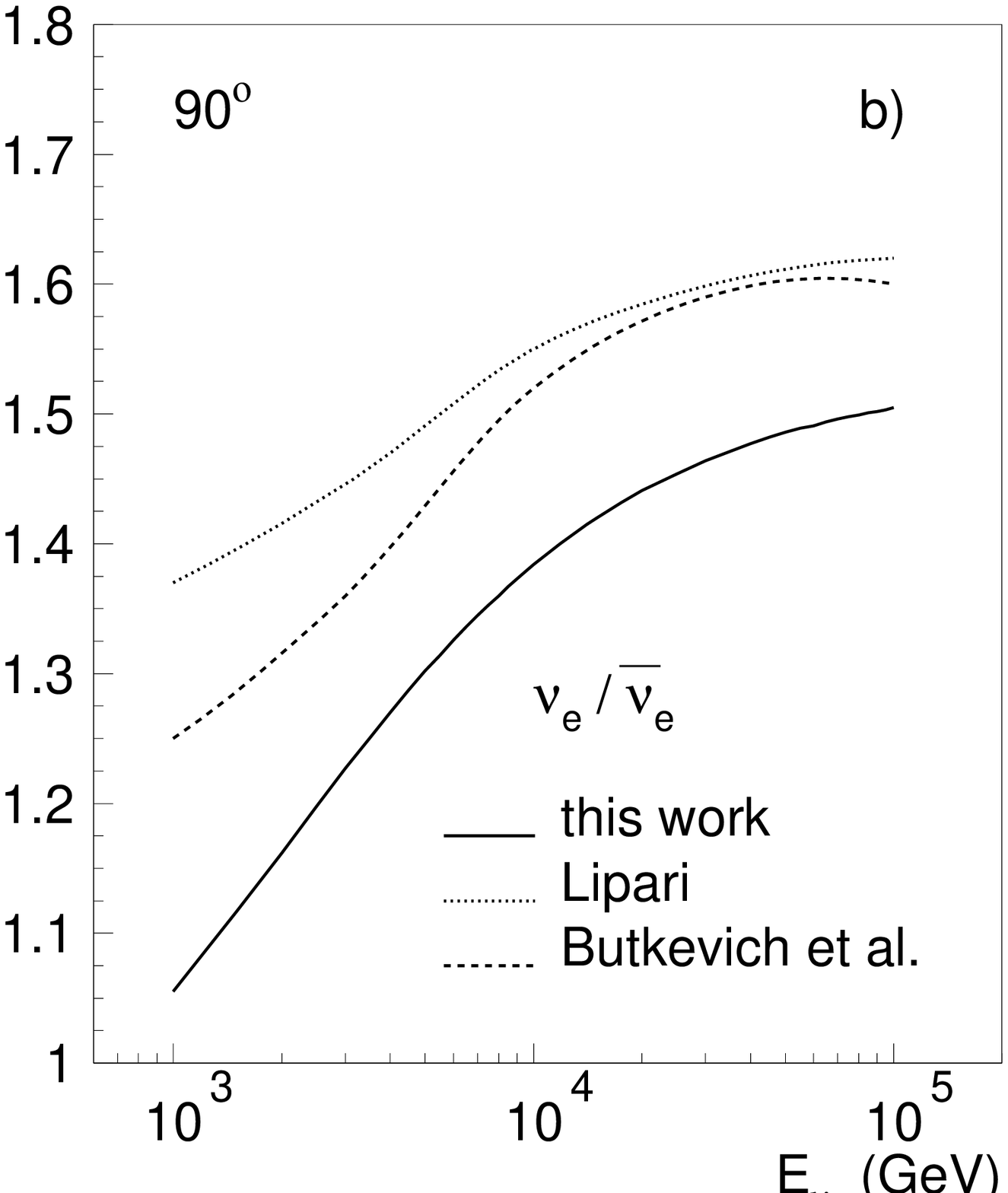,width=6.4cm,height=6.4cm}}
 \center\mbox{\epsfig{file=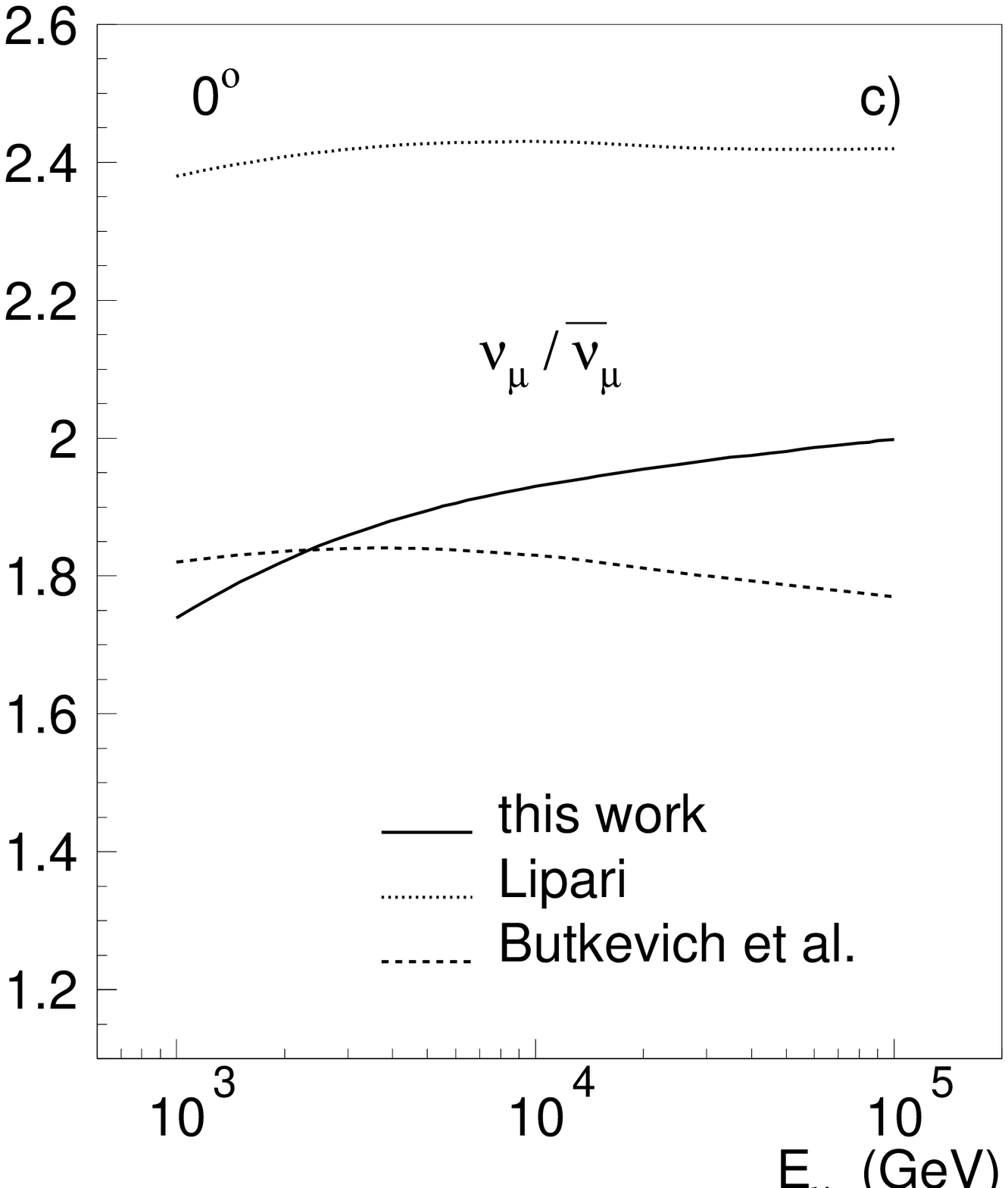,width=6.4cm,height=6.4cm}
     \hspace{2mm}\epsfig{file=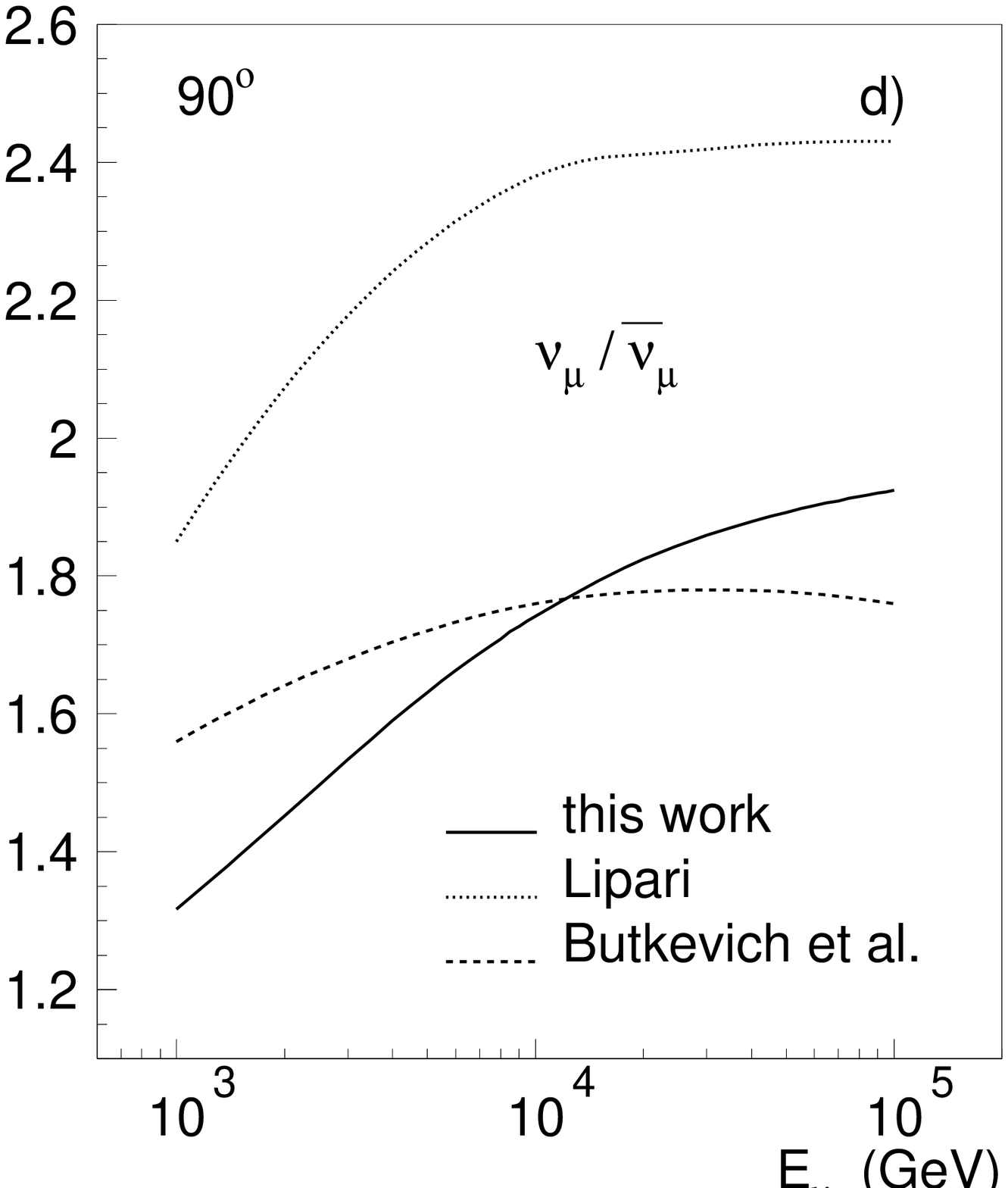,width=6.4cm,height=6.4cm}}
 \protect\caption{Neutrino to antineutrino ratios \emph{vs.} energy
                 at $\vartheta=0^\circ$ and $90^\circ$ for the
                 conventional AN flux in comparison with the results
                 by Lipari~\protect\cite{Lipari93} and Butkevich
                 et al.~\protect\cite{Butkevich89}.
 \label{f:6}}
 \end{figure}

\begin{figure}[htb]
 \center\mbox{\epsfig{file=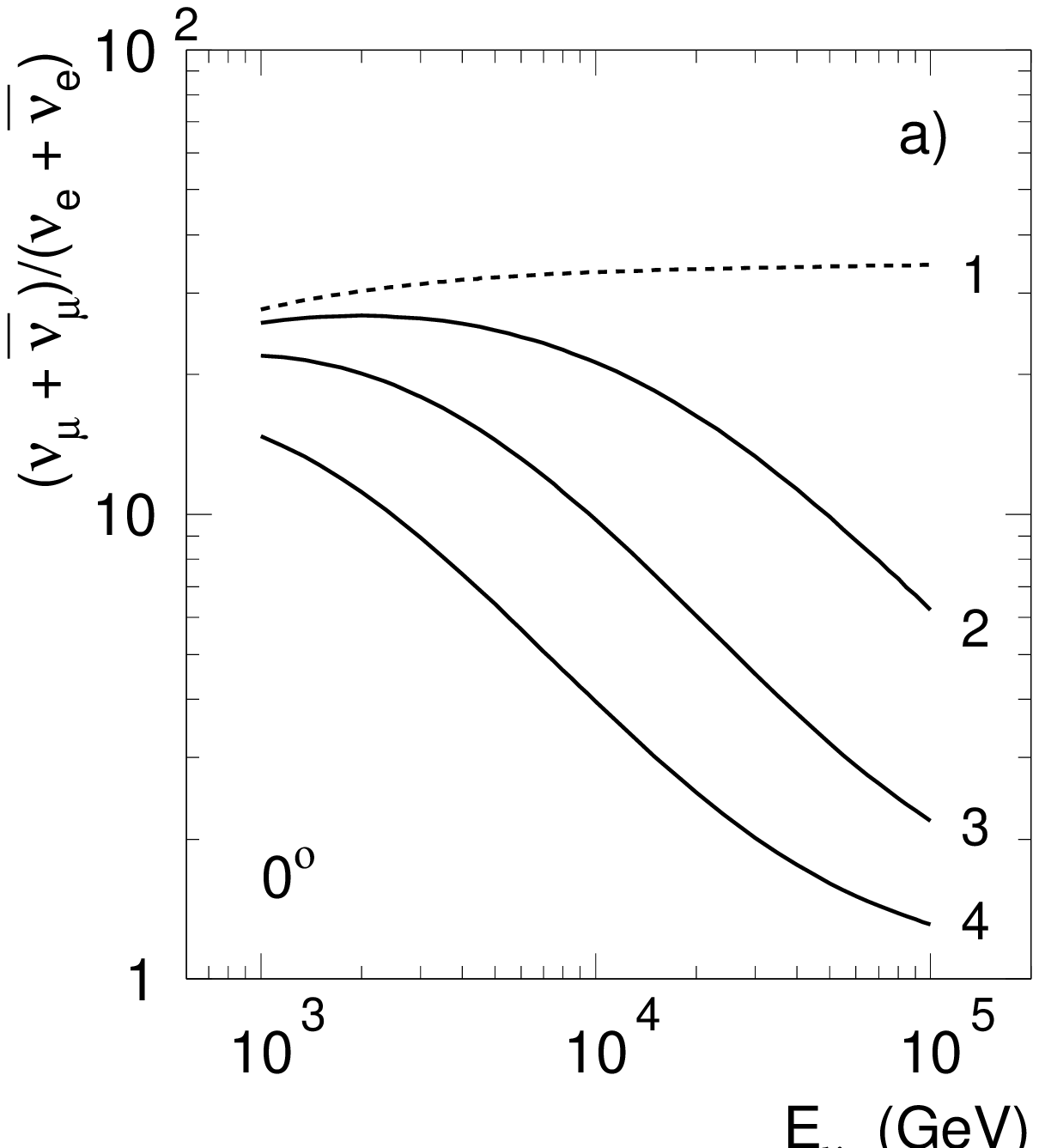,width=6.5cm,height=6.5cm}
               \epsfig{file=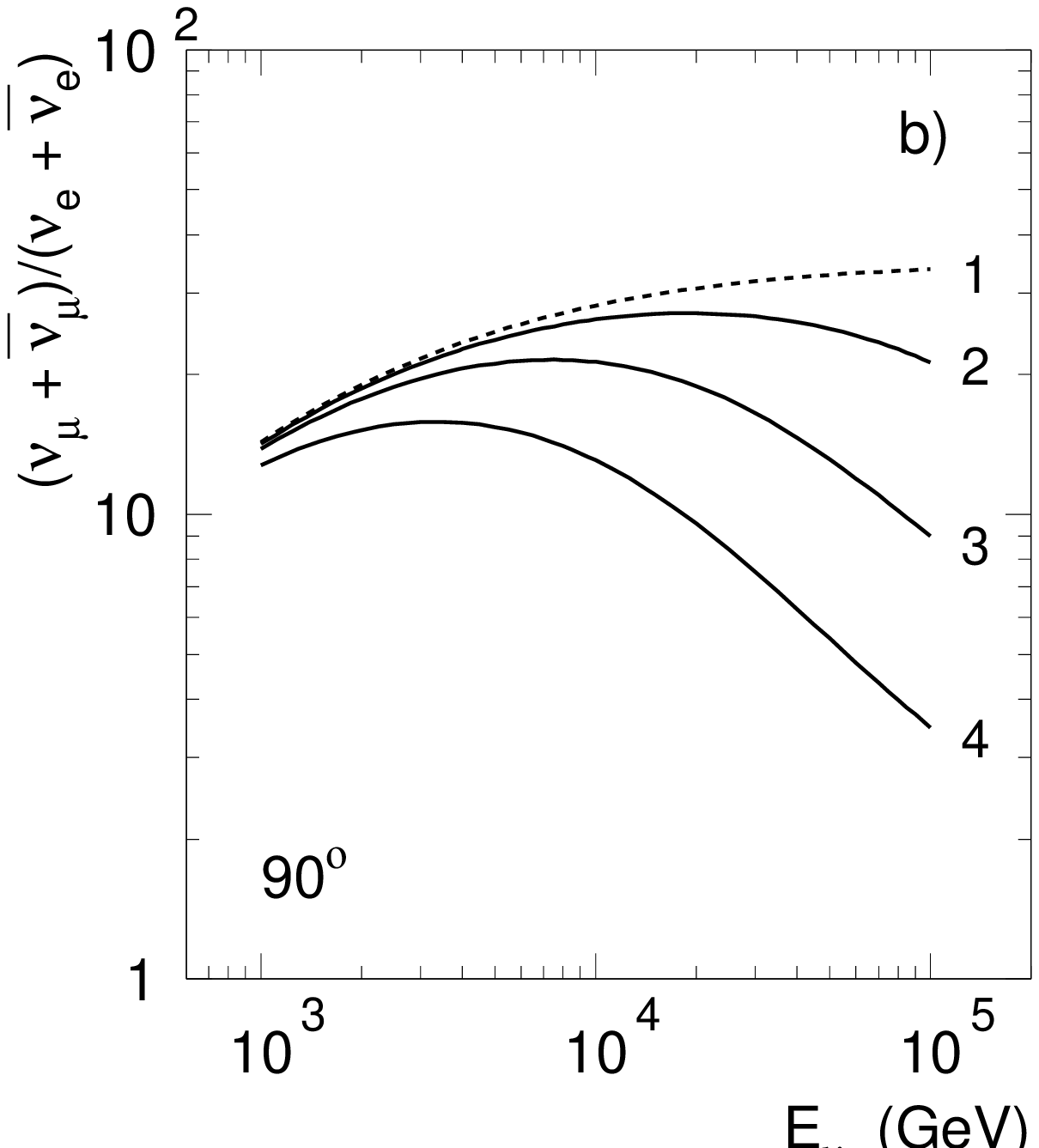,width=6.5cm,height=6.5cm}}
 \protect\caption{Neutrino flavor ratio, $R$, \emph{vs.} energy at
                  $\vartheta = 0^\circ$ (a) and $90^\circ$ (b) for
                  the total AN flux calculated without the PN
                  contribution (1) and with taking it into account
                  using the three models for charm production,
                  pQCD (2), QGSM (3) and RQPM (4).
 \label{f:7}}
 \end{figure}

\begin{figure}[htb]
 \center\mbox{\epsfig{file=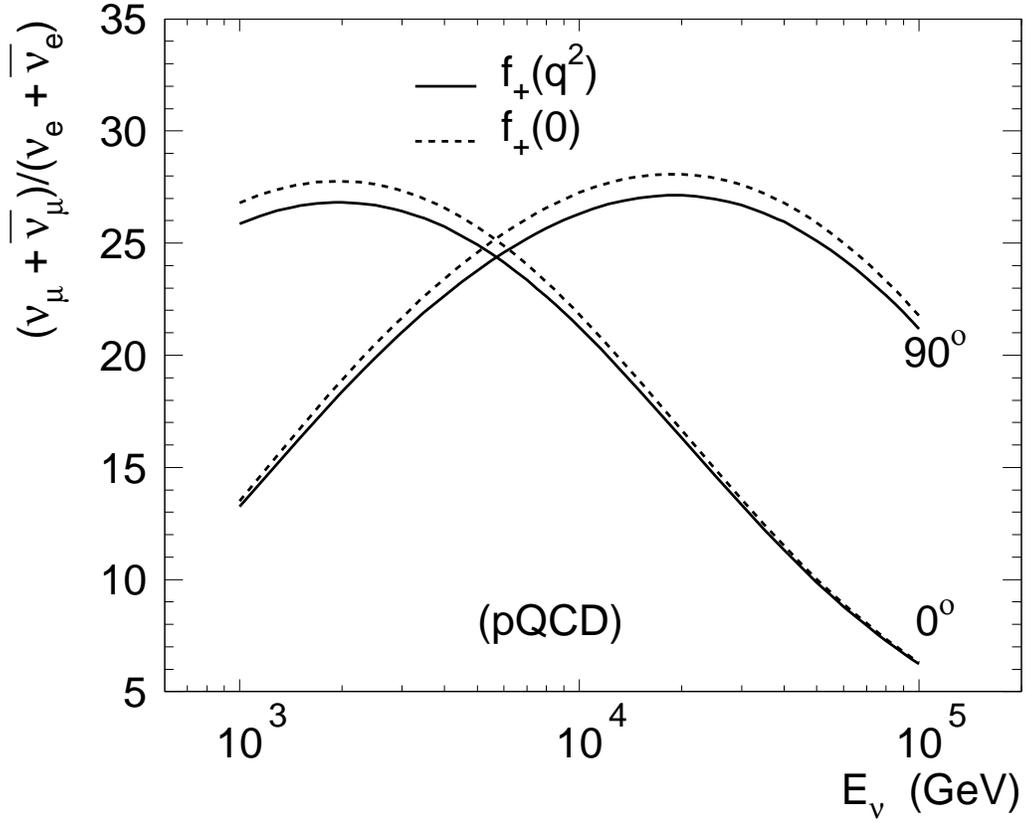,width=13.5cm}}
 \protect\caption{Effect of the $q^2$-dependent $K_{\ell 3}$ form
                 factors for the neutrino flavor ratio at
                 $\vartheta = 0^\circ$ and $90^\circ$.
                 The PN contribution is taking into account using the
                 pQCD model by Thunman et al.~\protect\cite{Thunman95}.
                 The dashed and solid curves are for the constant and
                 $q^2$-dependent form factors, respectively.
 \label{f:8}}
 \end{figure}

Let us discuss the numerical results presented in
figures~\ref{f:3}--\ref{f:8}.

Figures~\ref{f:3} and~\ref{f:4} show the individual contributions
from $\pi_{\mu2}$, $K^\pm_{\mu2}$, $K^0_{\ell 3}$, $K^\pm_{\ell 3}$
and $\mu^\pm_{e3}$ decays into the $\nu_e+\overline{\nu}_e$ and
$\nu_\mu+\overline{\nu}_\mu$ fluxes for vertical and horizontal
directions, as well as the overall $\pi,K$-neutrino fluxes. As an
example of possible PN contribution, we also show the results
obtained with the three alternative models for charm hadroproduction:
the recombination quark-parton model (RQPM), quark-gluon string model
(QGSM) and the model based on perturbative quantum chromodynamics
(pQCD). The basic assumptions of the first two models were described
by Bugaev et al.~\cite{Bugaev89} (see also \cite{Bugaev93} and references
therein). The third, ``state-of-the-art'' model was proposed recently by
Thunman et al.~\cite{Thunman95} to simulate charm hadroproduction
through pQCD processes. To leading order in the coupling constant,
$\alpha_s$, these are the gluon-gluon fusion and the quark-antiquark
annihilation. The next-to-leading order contributions are taken into
account by doubling the cross sections. To simulate the primary and
cascade interactions, the authors use the well-accepted Monte Carlo
code {\sl PYTHYA}.

It is our opinion that the RQPM and QGSM give the safe upper and lower
limits for the prompt muon and neutrino fluxes. These limits are not
inconsistent with the current deep underground measurements of the
muon intensity. However, considering rather strong discrepancy between
the data of the ground-based and underground muon
experiments~\cite{Bugaev93}, the comparatively low prompt muon
contribution predicted by the pQCD model cannot be excluded. Similar,
very low prompt muon contribution has been evaluated recently by
Battistoni et al.~\cite{Battistoni96} using the {\sl DPMJET-II}
code based on the two-component dual parton model and interfaced to
the shower code {\sl HEMAS}. The calculation of Battistoni
et al.~\cite{Battistoni96} does not yield the absolute prompt muon flux
but, from the estimated prompt-to-conventional muon ratio, one can see
a leastwise qualitative agreement with the result of the pQCD model by
Thunman et al.~\cite{Thunman95}. In particular, both models
predict that the prompt muon contribution overcomes the vertical
$\pi,K$-muon flux in the region of a thousand TeV and therefore is
undistinguished in the present-day muon experiments.

As may be seen from fig.~\ref{f:3}, the $K_{e3}$ decays give the
main contribution into the conventional $\nu_e+\overline{\nu}_e$ flux
above 1~TeV independent of zenith angle. The $K_{e3}^\pm$ and
$K_{e3}^0$ contributions are practically equal at $\vartheta=0^\circ$
and close in magnitude at $\vartheta=90^\circ$, despite the 8-fold
difference between the $K_{e3}^\pm$ and $K_{e3}^0$ decay branching
ratios. The first reason is that the life time of $K_L^0$ is about
4.2 times that of $K^\pm$. The second one lies in the different
inclusive cross sections for $K^++K^-$ and $K^0+\overline{K}{}^0$
production (see Vall et al.~\cite{Vall86}).
For the conventional $\nu_\mu+\overline{\nu}_\mu$ flux, the main
contribution comes from $K_{\mu2}^\pm$ decays (fig.~\ref{f:4}). At
$\vartheta = 90^\circ$ it stands out above the PN
contribution predicted by RQPM, up to about 200~TeV. Second in
importance is the $\pi_{\mu2}$ decay contribution. The cross
sections for pion production are almost order of magnitude larger
than for kaon production, but pion decays become rare above 1~TeV
owing to the large life time and Lorentz factor compared to kaon
ones. The $K_{\mu3}$ decays give comparatively small contribution
which however cannot be neglected.  Muon decay contribution is
negligible at $\vartheta = 0^\circ$ and very small at
$\vartheta = 90^\circ$.

In figures~\ref{f:5} (a--d) we present the conventional
$\nu_e+\overline{\nu}_e$ and $\nu_\mu+\overline{\nu}_\mu$
fluxes at $\vartheta = 0^\circ$ and $90^\circ$ as calculated by
Volkova~\cite{Volkova80}, Mitsui et al.~\cite{Mitsui86},
Butkevich et al.~\cite{Butkevich89}, Lipari~\cite{Lipari93}
and Agrawal et al.~\cite{Agrawal96}. All these fluxes are
normalized to our one.
In the multi-TeV energy range and above, our results fall within
the lowest and highest predictions. At $E_\nu < 2-3~TeV$ and
$E_\nu < 10~TeV$ for, respectively, $\vartheta = 0^\circ$ and
$\vartheta = 90^\circ$, our calculations give somewhat excessive
$\nu_\mu+\overline{\nu}_\mu$ fluxes. In part, this is due to our
simplified consideration of the pion decay effect (see
Section~\ref{sec:NC}). The corresponding discrepancy is almost
negligible for electron neutrinos, since pions give no direct
contribution to the $\nu_e+\overline{\nu}_e$ production.

In fig.~\ref{f:6} (a--d) we show our result for the conventional
neutrino to antineutrino ratios at $\vartheta=0^\circ$ and $90^\circ$
(\emph{vs.} neutrino energy) together with the results of Butkevich
et al.~\cite{Butkevich89} and Lipari~\cite{Lipari93}. The
$\nu_e/\overline{\nu}_e$ and $\nu_\mu/\overline{\nu}_\mu$ ratios are
very sensitive to the input parameters like the $n/p$ ratio, which is
governed by the primary chemical composition, and the $\pi^+/\pi^-$,
$K^+/K^-$, $K^\pm/K^0$ ratios, which are determined by the cross
sections for the meson production, regeneration and overcharging.
Whilst the parameters adopted by Butkevich et al.~\cite{Butkevich89},
Lipari~\cite{Lipari93} and in the present calculation are all within
the limits dictated by the available cosmic-ray and accelerator data,
the sets of the parameters differ considerably from each other.
Consequently, it is no great surprise that the predictions of
different models vary over a wide range.

On the other hand, owing to the well-known difference between
the cross sections for $\nu_\ell N$ and $\overline{\nu}_\ell N$
charged and neutral current induced interactions, the
$\nu_e/\overline{\nu}_e$ and $\nu_\mu/\overline{\nu}_\mu$ ratios
are very important inputs for the correct evaluating the
neutrino-induced throughgoing muon flux and contained event rate in
the neutrino detectors%
\footnote{At PeV energies, the $\nu_e/\overline{\nu}_e$ ratio
          becomes a crucial parameter on account of the $W$ resonance
          in $\overline{\nu}_e e^-$ annihilation.}.
As is seen from fig.~\ref{f:6}, the present-day uncertainty in the
$\nu_\mu/\overline{\nu}_\mu$ ratio is not satisfactory. The situation
with the $\nu_e/\overline{\nu}_e$ ratio proves to be somewhat better.

Figures~\ref{f:7} (a,b) show the neutrino flavor ratio, $R$,
\emph{vs} energy at $\vartheta=0^\circ$ and $90^\circ$. The dashed
curves represent the ratio for the conventional AN flux which is a
monotonically increasing function of energy varying from about 27.6
to 34.3 at $\vartheta=0^\circ$ and from about 13.3 to 33.6 at
$\vartheta=90^\circ$ within the interval $1 \div 100~TeV$. The
solid curves are for the $R$ evaluated with taking into account the
PN contribution according to three abovementioned models for charm
production. The PN contribution results in a decrease of the AN
flavor ratio, because the semileptonic decay modes of charmed hadrons
with $\nu_\mu$ ($\overline{\nu}_\mu$) and $\nu_e$ ($\overline{\nu}_e$)
in the final state have almost the same branching ratios.

As one can see, the $R$ is very sensitive to the charm production
model even at energies where the PN contribution remains small in
comparison with the conventional one. This effect provides an
interesting potential possibility for the experimental
discrimination of the PN production by measuring the ratio of the
``muonfull'' ($\nu_\mu$ and $\overline{\nu}_\mu$ induced) to
``muonless'' ($\nu_e$ and $\overline{\nu}_e$ induced) contained
event rates in a neutrino telescope for different energy thresholds
and directions. To a certain extent, such an experiment is easier
than the measurement of the absolute neutrino event rate.

Let us briefly sketch now the $K_{\ell3}$ form factor effect. Our
calculations show that the effect under discussion is almost
independent of zenith angle but quite different for muon and electron
neutrinos. At $E_\nu > 1~TeV$, the effect is identical for the
$K_{e3}^0$ and $K_{e3}^\pm$ contributions as well as for the overall
$\nu_e+\overline{\nu}_e$ flux (since the $K_{e3}$ decays are the main
source of $\nu_e$ and $\overline{\nu}_e$); its magnitude is just
higher than 3\,\%. The magnitude of the effect is different for
$K_{\mu3}^0$ (about 6\,\%) and $K_{\mu3}^\pm$ (about 4\,\%). But,
considering that the $K_{\mu3}$ contribution is by itself small,
there is no any change in the overall $\nu_\mu+\overline{\nu}_\mu$
flux. As a consequence, the inclusion of the $q^2$-dependent
$K_{\ell3}$ form factors decreases the
$(\nu_\mu+\overline{\nu}_\mu)/(\nu_e+\overline{\nu}_e)$ ratio for
the conventional neutrinos by about 3 to 4\,\%, depending on the
zenith angle and energy. As one might expect, the effect is small
but not completely negligible.
Figure~\ref{f:8} illustrates the $K_{\ell3}$ form factor effect for
the AN flavor ratio in the presence of a PN contribution. Clearly,
it is absent when prompt neutrinos dominate. Thus, we use the lowest
PM contribution as predicted in the pQCD model by
Thunman et al.~\cite{Thunman95}.
By the evident reasons, there is no effect of any value in the
$\nu_e/\overline{\nu}_e$ and $\nu_\mu/\overline{\nu}_\mu$ ratios.

\protect\section{Conclusions} \label{sec:Conc}

The main result of this work consists in the explicit formulas for
the semileptonic decay rates (differential and total) of $K^\pm$ and
$K^0_L$ mesons, which take into account the $q^2$-dependent
$K_{\ell3}$ form factors. The obtained rates differ essentially from
those calculated with constant form factors and the total rates are
in good agreement with experiment.

With a rather detailed model for nuclear cascade in the atmosphere
and with the improved differential $K_{\ell3}$ rates, we have
calculated the $\nu_e$, $\overline{\nu}_e$, $\nu_\mu$ and
$\overline{\nu}_\mu$ spectra at energies 1 to 100~TeV. The
calculated spectra are within the limits resulting from the
uncertainties in the current data on the primary cosmic ray flux and
cross sections for hadron-nucleus interactions at high energies.

The outcome of the inclusion of $q^2$-dependent $K_{\ell3}$ form
factors into the AN flux calculation is as follows.
  The electron neutrino flux from $K_{e3}^0$ and $K_{e3}^\pm$ decays
  increases by about 3--3.5\,\%.
  The $K_{\mu3}^0$ and $K_{\mu3}^\pm$ decay contributions into the
  muon neutrino flux reduces by about 6 and 4\,\%,
  respectively, whereas the change in the overall
  $\nu_\mu+\overline{\nu}_\mu$ flux is completely negligible.
  If the PN contribution is as slight as predicted by the pQCD model,
  the change in the neutrino flavor ratio, $R$, comprises 3 to 4\,\%.
  This small systematic effect may only slightly be enhanced by taking
  account of kaon regeneration or through the variation of the kaon
  production cross sections within the experimental boundaries. And
  vise versa, it may be removed, wholly or in part, beyond the
  multi-TeV energy range if the PN contribution is as large as it
  follows from the RQPM or QGSM predictions.

The sensitivity of the $R$ to the PN contribution provides a way
for an experimental discrimination of this contribution by measuring
the relationship between the  ``muonfull'' and ``muonless'' contained
event rates in a neutrino telescope.

\acknowledgments

This work was supported in part by the Ministry of General and
Professional Education of Russian Federation, project No.~728
within the framework of Program ``Universities of Russia --
Basic Researches''. V.~N. thanks INFN, Sezione di Firenze for its
hospitality.

%%%%%%%%%%%%%%%%%%%%%%%%% Figures %%%%%%%%%%%%%%%%%%%%%%%%%%%%

\end{document}